%% file: KNN_DBSCAN.tex
\documentclass[nonacm]{acmart}

\input preamble.tex

\input macros.tex

\AtBeginDocument{%
  }

\begin{document}


\title{KNN-DBSCAN: a DBSCAN in high dimensions}
\authorsaddresses{}

\author{Youguang Chen}
\email{youguang@utexas.edu}
\affiliation{%
  \institution{University of Texas at Austin}
  \department{Oden Institute for Computational Engineering and Sciences}
  \streetaddress{201 E. 24th Street, POB 4.102}
  \city{Austin}
  \state{Texas}
    \country{USA}
}

\author{William Ruys}
\email{will@oden.utexas.edu}
\affiliation{%
 \institution{University of Texas at Austin}
 \department{Oden Institute for Computational Sciences}
  \streetaddress{201 E. 24th Street, POB 4.102}
  \city{Austin}
  \state{Texas}
    \country{USA}
}

\author{George Biros}
\email{gbiros@acm.org}
\affiliation{%
  \institution{University of Texas at Austin}
 \department{Oden Institute for Computational Sciences}
   \streetaddress{201 E. 24th Street, POB 4.102}
  \city{Austin}
  \state{Texas}
   \country{USA}
}


\begin{abstract} 
\input{s0_abstract}
\end{abstract}




\maketitle

\section{Introduction} \label{s:intro} \input s1_intro.tex
\section{\textit{k}NN-DBSCAN} \label{s:methods}
\input s2_methods.tex
\input s2s_theory.tex

\section{Parallel Implementation of \textit{k}NN-DBSCAN} \label{s:algorithms} 
\input s2s_algos.tex

\section{Experiments} \label{s:results}

\input s3_results.tex

\section{Conclusions}\label{s:conclusionss} \input s4_conclusions.tex

\bibliographystyle{ACM-Reference-Format}
\bibliography{refs}

\appendix
\input appendix/proofs.tex

\input appendix/experiment_additional.tex

\end{document}

%% file: preamble.tex
\usepackage{graphics,epsfig,graphicx,float,color}
\usepackage{fancyhdr}
\usepackage{textcomp}
\usepackage{algorithm}

\usepackage[noend]{algpseudocode}

\usepackage{verbatimbox}
\usepackage{etex}
\usepackage{amsmath,amsbsy,amsfonts,latexsym}
\usepackage{multicol,multirow}
\usepackage{soul}
\usepackage{hhline}
\usepackage{paralist}
\usepackage{array}
\usepackage{marvosym}
\usepackage{url}
\usepackage{boxedminipage}
\usepackage{multirow}
\usepackage{threeparttable}
\usepackage{pdfpages}
\usepackage{xspace}
\usepackage{colortbl}
\usepackage{tabularx,booktabs}
\usepackage{xcolor}
\usepackage{tikz} 
\usetikzlibrary{arrows} 
\usetikzlibrary{patterns}  
\usepackage{pgfplots} 
\pgfplotsset{compat=1.3}

\usepackage{enumitem}
\usepackage{siunitx}
\usepackage{booktabs}
\usepackage{calligra}

\DeclareMathAlphabet{\mathcalligra}{T1}{calligra}{m}{n}

\definecolor{light-gray}{gray}{0.8}

\newcolumntype{R}{>{\columncolor{light-gray}}r}
\newcolumntype{L}{>{\columncolor{light-gray}}l}
\newcolumntype{C}{>{\columncolor{light-gray}}c}

\tikzset{zigzag/.style={decorate, decoration=zigzag}}

\renewcommand{\arraystretch}{1.2}

\newtheorem{definition}{Definition}[section]

\usepackage{mathtools}
\DeclareMathSymbol{\mh}{\mathord}{operators}{`\-}

\DeclareMathOperator{\directRadireach}{\xrightarrow{\epsilon}}
\DeclareMathOperator{\radireach}{\xrightharpoonup{\epsilon}}
\DeclareMathOperator{\radicon}{\xleftrightarrow{\epsilon}}

\DeclareMathOperator{\core}{\mathcal{I}_{core}}
\DeclareMathOperator{\border}{\mathcal{I}_{border}}
\DeclareMathOperator{\noise}{\mathcal{I}_{noise}}
\DeclareMathOperator{\Neps}{\mathcal{N}_\epsilon}

\algnewcommand\algorithmicforeach{\textbf{for each}}
\algdef{S}[FOR]{ForEach}[1]{\algorithmicforeach\ #1\ \algorithmicdo}
\algdef{SE}[DOWHILE]{Do}{doWhile}{\algorithmicdo}[1]{\algorithmicwhile\ #1}%
\usetikzlibrary{matrix}
\usetikzlibrary{arrows}

\newcommand{\commentsymbol}{//}
\algrenewcommand\algorithmiccomment[1]{\hfill \commentsymbol{} #1}
\algnewcommand{\LineComment}[1]{\State \(\commentsymbol{}\) #1}
\usepackage{setspace}
\usetikzlibrary{positioning}

\definecolor{t_green}{rgb}{0.533333333,	0.847058824,	0.690196078}
\definecolor{t_yellow}{rgb}{1,	0.964705882,	0}
\definecolor{t_orange}{rgb}{1,	0.8,	0.360784314}
\definecolor{t_blue}{rgb}{0.607843137,	0.866666667,	1}
\definecolor{t_red}{rgb}{1,0.435294118,0.411764706}

\usepackage{times,helvet,courier,xcolor}
\usepackage{tabularx,ragged2e,booktabs,caption}
\newcolumntype{C}{>{\Centering\arraybackslash}X} %
\usepackage{textcmds}

\usetikzlibrary[pgfplots.groupplots,calc]

\pgfplotsset{grid style={dotted,black}}
\pgfplotsset{major grid style={dotted,black}}

\usepackage{nameref}
\usepackage{cleveref}
\usepackage{subcaption}
\newcounter{mylabelcounter}

\makeatletter
\newcommand{\labelText}[2]{%
#1\refstepcounter{mylabelcounter}%
\immediate\write\@auxout{%
  \string\newlabel{#2}{{1}{\thepage}{{\unexpanded{#1}}}{mylabelcounter.\number\value{mylabelcounter}}{}}%
}%
}
\makeatother

\usepackage{siunitx,paralist,pifont}

%% file: macros.tex
\newcommand{\enn}{$\epsilon$-NNG}
\newcommand{\knn}{$k$-NNG}
\newcommand{\dbsc}{$k$NN-DBSCAN}

\usepackage{xparse}
\NewDocumentCommand{\var}{O{s} m O{}}{%
  \ensuremath{#1_{#2}^{#3}}
}

\newcommand{\figref}[1]{Figure~\ref{#1}}

\newcommand{\secref}[1]{\S\ref{#1}}

\newcommand{\defref}[1]{Definition~\ref{#1}}
\newcommand{\lemmaref}[1]{Lemma~\ref{#1}}

\DeclareMathOperator*{\argmin}{arg\,min}
\DeclareMathOperator{\bigO}{\mathcal{O}}

\newcommand{\ipoint}[1]{\textit{\textbf{\color{darkgray}#1}}}

\newcommand{\idef}[1]{\textit{\textbf{#1}}}

\newcommand{\MA}[1]{\ensuremath{\mathcal{#1}}}

\newcommand{\tn}{\widetilde{n}}

\newcommand{\bipa}{\begin{inparaenum}[(\itshape i\upshape)]}
\newcommand{\eipa}{\end{inparaenum}}
\newcommand{\bipasub}{\begin{inparaenum}[(\itshape a\upshape)]}
\newcommand{\eipasub}{\end{inparaenum}}

\definecolor{light-gray}{gray}{0.80}
\definecolor{bittersweet}{rgb}{1.0, 0.44, 0.37}
\definecolor{darkgreen}{rgb}{0.1, 0.75, 0.24}
\definecolor{darkmagenta}{rgb}{0.55, 0.0, 0.55}
\definecolor{darkpink}{rgb}{0.94, 0.19, 0.22}
\definecolor{ferrarired}{rgb}{1.0, 0.11, 0.0}
\definecolor{oxfordblue}{rgb}{0.2, 0.8, 0.8}

\renewcommand\paragraph{\subsubsection*}

\protected\def\vtsp{%
  \ifmmode
    \mskip0.5\thinmuskip
  \else
    \ifhmode
      \kern0.08334em
    \fi
  \fi
}

%% file: s0_abstract.tex
Clustering is a fundamental task in machine learning. One of the most successful and broadly used algorithms is DBSCAN, a density-based clustering algorithm. DBSCAN requires $\epsilon$-nearest neighbor graphs of the input dataset, which are computed with range-search algorithms and spatial data structures like KD-trees. Despite many efforts to design scalable implementations for DBSCAN, existing work is limited to low-dimensional datasets, as constructing $\epsilon$-nearest neighbor graphs can be expensive in high-dimensions. This paper introduces a modified DBSCAN, using $k$-nearest neighbor ($k$NN) graphs to improve efficiency. We outline conditions for $k$NN-DBSCAN to match DBSCAN's results and present a parallel implementation using OpenMP and MPI for shared and distributed memory systems. Testing on datasets up to 32 dimensions, we achieve remarkable scalability. Our implementation clusters one billion 3D points in under one second on 28K cores at TACC's Frontera system. In a larger run, we cluster 65 billion points in 20 dimensions in under 40 seconds using 114,688 cores. Our method is up to 37$\times$ faster than state-of-the-art parallel DBSCAN on a 20-dimensional dataset with 4 million points. Code is available at \url{https://github.com/ut-padas/knndbscan}.

%% file: s1_intro.tex
\begin{figure}[t]
\centering
\begin{tikzpicture}
    \node[inner sep=0pt] (a) at (0,0) {\includegraphics[width=12cm]{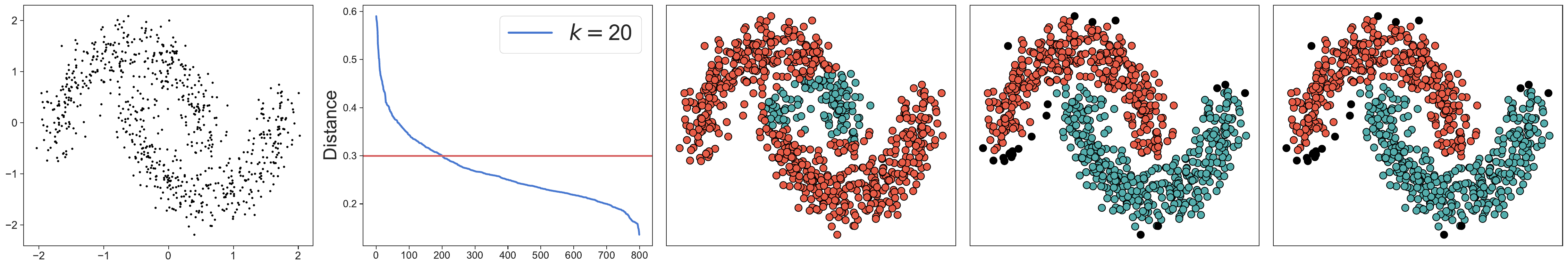}};
    \node[inner sep=0pt] (b) at (0,-2) {\includegraphics[width=12cm]{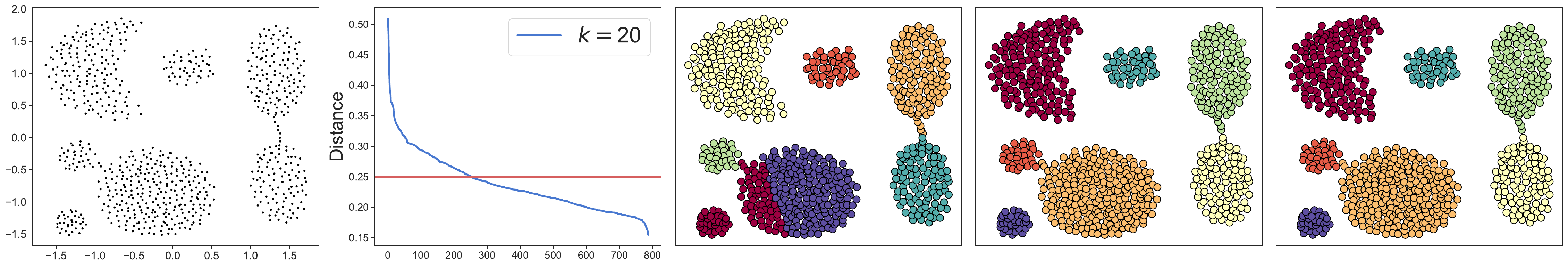}};
    \node[inner sep=0pt] (c) at (0,-4) {\includegraphics[width=12cm]{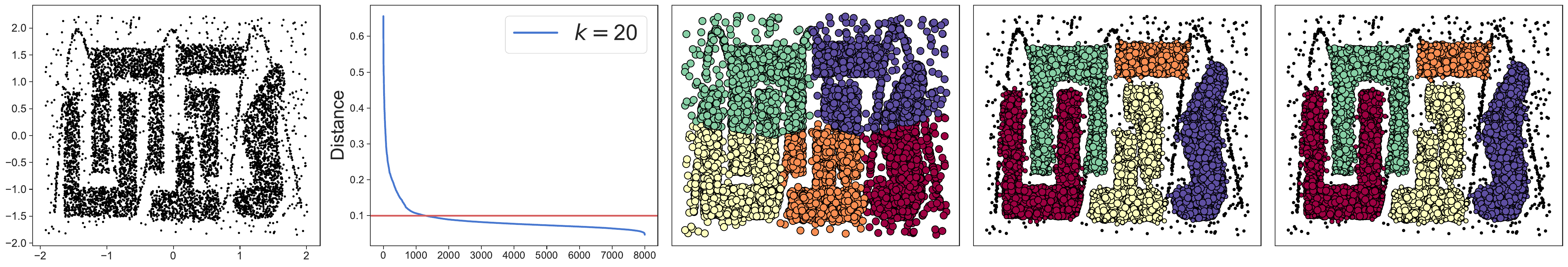}};
    \node[inner sep=0pt] (d) at (0,-6) {\includegraphics[width=12cm]{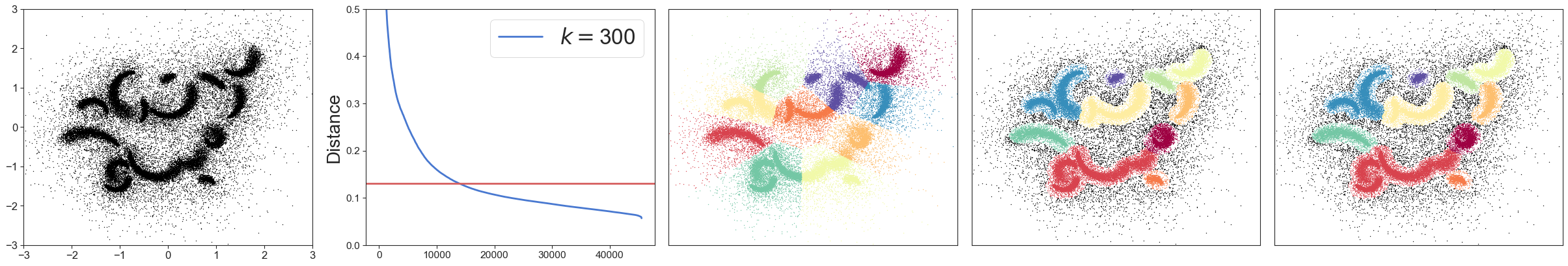}};
    \node[]  at (-2.2,1.2) {\small $k$-distance plot};
    \node[]  at (0.2,1.2) {\small K-Means};
    \node[]  at (2.5,1.2) {\small DBSCAN};
    \node[]  at (5,1.2) {\small \dbsc{}};
\end{tikzpicture}
\caption{\color{black}{Clustering results using K-Means, DBSCAN, and \dbsc{} on 2D datasets. The second column shows the sorted $k$-distance plot for each dataset. Both DBSCAN and \dbsc{} require two parameters: the range $\epsilon$ and the minimum number of points needed to form a dense region ($minPts$). For both algorithms, we set $minPts$ to the value of $k$ in the $k$-distance plots, with $\epsilon$ indicated by the red line in the plots. The \dbsc{} algorithm produces \textbf{identical} clustering results to DBSCAN in these examples.}}
\label{fig:2D}
\end{figure}

Given a set $\MA{I}$ of $n$ points,  $\{p_i\}_{i=1}^n$,  $p_i \in \mathbb{R}^d$,  we seek a mapping from $\MA{I}$ to $\MA{C}:=\{1,\ldots,c\}$ that groups $\MA{I}$ in $c$ different clusters. The dimensionality $d$ of the dataset plays a critical role in determining the difficulty of the clustering task. \idef{Density-Based Spatial Clustering of Applications with Noise}  (DBSCAN)~\cite{ester1996dbscan}  is one  popular and practical algorithms for clustering in metric spaces. Unlike the iterative $k$-means~\cite{gan2017dbscan} algorithm, DBSCAN does not require initialization and comprises only two hyperparameters. Moreover, it possesses the ability to automatically detect and disregard noisy data points within the dataset. DBSCAN's effectiveness extends beyond that of simpler algorithms like $k$-means~\cite{jain10}, as it can cluster non-linearly separable points, as demonstrated in \figref{fig:2D}. However, it is important to note that DBSCAN is more intricate, computationally more expensive, and less scalable than $k$-means. While $k$-means can be easily implemented and scales as $\bigO(d\vtsp c\vtsp n/p + \log p)$, with $p$ being the number of MPI processes~\cite{dhillon2002kmeans}, existing DBSCAN implementations may exhibit a complexity of up to $\bigO(d \vtsp n^2 /p)$ for datasets with large dimensions ($d$)~\cite{weber-schek-blott98,indyk-andoni08}.

\color{black}
A significant amount of research has focused on developing parallel DBSCAN algorithms \cite{patwary2012,patwary2013,patwary-dubey-e14,patwary-dubey-e15,griddbscan,sarma2019mudbscan}. One notable limitation of these existing parallel DBSCAN algorithms is their scalability with high-dimensional datasets. For instance, Patwary et al.~\cite{patwary2012} introduced PDBSCAN, which constructs parallel DBSCAN by creating a weighted graph known as the \idef{$\epsilon$-Nearest-Neighbor Graph} (\enn{}) from the input points.\color{black} While tree data structures~\cite{samet06} can efficiently construct \enn{} in low dimensions with a complexity of $\bigO(d n\log n)$, this complexity escalates to $\bigO(d n^2)$ in high-dimensional spaces~\cite{andoni2006near,indyk-andoni08}. Additionally, the \enn{} structure becomes increasingly sensitive to changes in the $\epsilon$ parameter as dimensionality increases.\color{black} Kumari et al.~\cite{griddbscan} and Wang et al.~\cite{wang2019dbscan} suggest using grids to divide the data space into smaller cells and optimizing the neighborhood query by restricting the search to neighboring cells. However, the number of cells grows exponentially with dimensionality, which adversely affects the method's performance on high-dimensional data.

Another limitation of existing parallel DBSCAN algorithms is their inefficiency in finding optimal input hyperparameters, $\epsilon$ and $minPts$. Properly setting these parameters is crucial for obtaining meaningful clustering results. One heuristic approach for selecting parameters involves first determining $minPts$, as recommended by the original DBSCAN paper~\cite{ester1996dbscan} and \cite{schubert-2017dbscan}. This value helps smooth the density estimate. Next, we find the $minPts$-nearest neighbors for each point and plot the sorted distances to these neighbors. The appropriate $\epsilon$ is often selected at the "elbow" or "knee" of this plot, as illustrated in \Cref{fig:2D}. This process requires adaptively searching for suitable parameter candidates and running the clustering algorithm multiple times. However, with existing parallel DBSCAN algorithms, any adjustment to the parameters $\epsilon$ or $minPts$ requires restarting the algorithm and performing the range query from scratch again. This limitation hinders the application of DBSCAN for large-scale data, where a single run can be time-consuming.

To address these two limitations, we introduce \dbsc{}, which is based on but differs from DBSCAN. \dbsc{} shares the same hyperparameters and definitions for core and noise points as DBSCAN, but it uses an alternative reachable condition to define clusters. Details and justifications for this new method are provided in \Cref{s:methods}. The advantage of \dbsc{} is its ability to leverage the \idef{$k$-Nearest-Neighbor graph} ($k$-NNG). \color{black} By employing a randomized algorithm for constructing $k$-NNG, \dbsc{} offers improved scalability and speed compared to $\epsilon$-NNG, particularly for large-scale or high-dimensional data. For instance, our implementation uses the open-source GOFMM library~\cite{gofmm-home-page,yu-reiz-biros18} to construct an approximate $k$-NNG. GOFMM utilizes a randomized projection tree algorithm~\cite{xiao2016parallel}, which performs based on the dataset's \emph{intrinsic dimensionality} rather than its ambient dimensionality~{\cite{intrinsic}}. When the intrinsic dimension is low, the complexity of randomized projection trees converges asymptotically to $\bigO(d n \log n)$~\cite{jones2011randomized}. However, if the intrinsic dimension is high, one can either manage the accuracy or the computational complexity—but not both; in such cases, the complexity may reach $\bigO(d n^2)$. The memory requirements remain $\bigO(n k)$.

\color{black}
Another advantage of \dbsc{} is that after constructing the $k$-NNG, the algorithm can efficiently handle any combination of input parameters $\epsilon$ and $minPt \leq k$ without the need to re-query neighbors from the beginning. This method can be considerably more efficient in practice compared to existing parallel DBSCAN implementations, as constructing $k$-NNG (or $\epsilon$-NNG for DBSCAN) generally takes more time than the clustering itself. For instance, in our implementation, building the $k$-NNG for a 64 million 5D dataset with $k=100$ takes 400 times longer than the clustering process (see \Cref{s:scaling} for details).

\color{black}
Our  parallel implementation of \dbsc{} uses a lock-free shared memory algorithm (in OpenMP) and MPI-based distributed algorithm. To minimize communication costs, we also use an approximate MST algorithm (as opposed to the exact MST of the graph). We show that this approximation doesn't change the clustering quality in \secref{s:inexact-mst}. 

\color{black}
The main contributions of the paper are as follows:
\begin{itemize}[leftmargin=*]
    \item We propose a novel density-based clustering algorithm, \dbsc{}, which is inspired by but distinct from DBSCAN, as discussed in \Cref{s:knn-dbscan}. \dbsc{} enables the use of $k$-NNG for clustering. We provide a theoretical justification that \dbsc{} is well-defined and establish the relationships between \dbsc{} and DBSCAN when using the same input parameters (see \Cref{thm:relation-to-dbscan}).
    \item We introduce an inexact-MST method to produce clustering results for both DBSCAN and \dbsc{}, designed to be compatible with distributed memory architectures (see \Cref{s:inexact-mst}). We prove that this approximation does not compromise clustering quality in \Cref{thm:inexactmst}.
    \item We propose a hybrid MPI/OpenMP implementation of \dbsc{} that uses Boruvka's algorithm to construct the MST (see \Cref{s:algorithms}). Empirical results in \Cref{s:acc} demonstrate that \dbsc{} performs comparably to, and occasionally better than, DBSCAN on small-scale datasets. In \Cref{s:scaling}, we present strong and weak scaling results for our parallel \dbsc{} implementation using synthetic datasets with up to 65 billion points and 20 dimensions. Additionally, we compare our implementation with the parallel DBSCAN algorithm PDBSCAN~\cite{patwary2012} on real-world datasets (two MNIST variants and CIFAR10) in \Cref{s:comparison}, showing that our approach outperforms it.
\end{itemize}

\color{black}

\color{black}
\section{Related work}
\color{black}

\textbf{DBSCAN.} {Clustering is a key algorithm in machine learning and data analysis~\cite{jain10}. DBSCAN~\cite{ester1996dbscan} published at the KDD'96 data mining conference is a popular clustering algorithms since it is a deterministic, non-iterative algorithm, works on nonlinearly separable datasets, produces good quality clusters (nearly as good as the much more expensive spectral clustering), is robust to outliers, and has a minimum number of parameters. It is an algorithm proven to work in practice, and in 2014, it received the SIGKDD test-of-time award. For these reasons most scientific computing/machine learning frameworks like MATLAB, ELKI~\cite{ELKI}, R~\cite{R}, Weka~\cite{weka}, scikit-learn~\cite{scikit-learn} offer default DBSCAN implementations.}

Jiang et al.~\cite{jiang2017dbscan} present theoretical analysis of the performance and convergence of DBSCAN with increasing $n$. Its convergence is characterized in terms of the geometry of the level set of  the underlying probability density function of $\MA{I}$. It is shown that DBSCAN converges as $\bigO\left(n^{-1/r}\right)$, where $r$ here is the intrinsic dimension. The author also proposes concrete ways for choosing the hyperparameters. In \cite{jiang2020dbscan}, the authors proposes SNG-DBSCAN that runs in $\bigO\left( n\log n\right)$ time. The method reduces the complexity mainly by subsampling the edges of the neighborhood graph. The authors also show that the sampled version of DBSCAN can recover true clusters with statistical guarantees based on some assumptions.

\textbf{Parallel DBSCAN.} Patwary et al.~\cite{patwary2012} present PDBSCAN, a scalable DBSCAN implementation that uses MPI. It uses a KD-tree to construct the exact \enn{}.  The paper has a nice review of the DBSCAN literature prior to 2012. PDBSCAN takes an hour for a problem with 115M points in 10 dimensions on  8,192 AMD cores. PDBSCAN is publically available~\cite{pdbscan-patwary12} and we compare our implementation with it in~\secref{s:results}.  The same group proposes OPTICS, a DBSCAN-like clustering algorithm~\cite{patwary2013} that achieves similar performance.  However, due to its reliance on constructing the exact \enn{}, PDBSCAN faces scalability challenges with high-dimensional data and memory issues with datasets exhibiting varying densities.

Patwary et al.~\cite{patwary-dubey-e14} introduce PARDICLE, an OpenMP-plus-MPI algorithm, that scaled up to 4,096 MPI processes.  The key innovation is the use of an approximate $\epsilon$-graph by estimating the point density and restricting exact neighborhood searches to a small number of points.  This approximation is heuristic and not connected to the intrinsic dimensionality of the dataset.  The main limitation of PARDICLE is that $\epsilon$-searches (whenever needed) are done  using FLANN~\cite{muja-low-flann-14}. Since FLANN doesn't support distributed-memory parallelism, the algorithm in~\cite{patwary-dubey-e14}  requires an $\epsilon$-overlap  halo of the point set in each MPI rank.  In high-dimensions computing this halo is not scalable neither in terms of memory nor computation because it may require the whole dataset to be replicated in every MPI process.  The largest problem solved with PARDICLE was 115M points in 10 (ambient) dimensions (intrinsic dimension is unknown). Clustering  on 512 Xeon cores takes roughly 84 seconds.  
\color{black}

Kumari et al.~\cite{griddbscan} introduce GridDBSCAN, which partitions the data space into smaller cells and optimizes neighborhood queries by restricting the search to adjacent cells. The authors report that this approach can reduce neighborhood queries by up to 15\%. However, the number of cells increases exponentially with the dimension, which negatively impacts the method's performance on high-dimensional data. Sarma et al.~\cite{sarma2019mudbscan} present $\mu$DBSCAN, a method that reduces the number of neighborhood queries by exploiting spatial locality. They also offer an MPI-based implementation, which in one test on 1 billion points in 3D, took 41 minutes on 32 nodes (with 1 core per node). However, the algorithm's performance heavily relies on the assumption that the data exhibits strong spatial locality. In datasets where spatial locality is weak or non-existent, the efficiency gains from the $\mu$DBSCAN approach may be reduced. Especially in high-dimensional spaces, concept of spatial locality becomes less meaningful (curse of dimensionality).

Existing parallel DBSCAN algorithms have several limitations: (1) The scalability of these algorithms is challenged by high-dimensional datasets. (2) The $minPts$ parameter used in experiments is typically very low; for instance, \cite{patwary2012} uses a maximum $minPts$ of 20, while \cite{patwary-dubey-e14} uses 5, and \cite{griddbscan,sarma2019mudbscan} use 6. (3) If the DBSCAN input parameters $\epsilon$ or $minPts$ are adjusted, the algorithm have to be rerun from scratch. This is inefficient in practical applications, as determining the optimal parameters often requires running DBSCAN multiple times.

%


\color{black}

%% file: s2_methods.tex
In this section, we introduce our proposed \dbsc{} method. We start by reviewing the basic definition of DBSCAN in \secref{s:dbscan}, then we define \dbsc{} in \secref{s:knn-dbscan}. In \secref{s:mst}, we explain the relationships between MST, DBSCAN, and \dbsc{}. Next, we introduce an approximate version of MST in \secref{s:inexact-mst}, which is used for the parallel implementation of \dbsc{} in \secref{s:algorithms}.

%% file: s2s_theory.tex
\begin{table}[t!]
\footnotesize
    \centering
    \caption{Summary of notation.}
    \label{notation}
    \begin{tabularx}{\columnwidth}{>{\hsize=0.25\hsize}X
                              >{\hsize=0.7\hsize}X}
        \toprule
        Notation     & Description     \\ \midrule
        $\mathcal{I}$         & the entire set of points       \\
        $k_{\max}$          &the number of nearest neighbor points in \knn{} \\ 
        $\core, \border$, $\noise$       & core points, border points and noise points       \\ 
        
        $\epsilon$          &\ipoint {the radius of a neighborhood}    \\ 
        $M$          &\ipoint{the minimum neighborhood size of core points}   \\ 
        $d(p,q)$     &the Euclidean distance between point $p$ and $q$\\ 
        $\Neps(p)$          &the $\epsilon\mh$radius neighborhood of point p  \\ 
        $\mathcal{N}_k(p)$          &the $k\mh$nearest neighborhood of point p  \\ 
        $G=(\mathcal{I}, E)$          &a graph of points $\mathcal{I}$ and edges $E$  \\ 
        $G_\epsilon, \, G_{\epsilon,\text{core}}$          &the $\epsilon\mh$radius neighbor graph and its subgraph associated with core points\\ 
        $G_k, \, G_{k,\text{core}}$          &the $k\mh$nearest neighbor graph and its subgraph associated with core points\\ 
        $\directRadireach, \, \radireach \, \radicon$          &direct $\epsilon\mh$reachable,  $\epsilon\mh$reachable and $\epsilon\mh$connected\\
        $\xrightarrow{M}, \xrightharpoonup{M}, \xleftrightarrow{M}$          &direct $M\mh$reachable, $M\mh$reachable and $M\mh$connected\\ \bottomrule
    \end{tabularx}
\end{table}

\subsection{DBSCAN}\label{s:dbscan}
DBSCAN detects high-density spatial regions and expands them to form clusters. \idef{It has two hyperparameters: the radius of neighborhood $\epsilon$ and the minimum number of points $M$} (also found as {\it minPts} in the literature). Using these parameters points are classified into three different types in Definition \ref{def:points} based on density, i.e., core points, border points and noise points.

\begin{definition}[classification of points]
\label{def:points}
Given $\epsilon$ and $M$, DBSCAN classifies points $\mathcal{I}$ into three types:
    \begin{itemize}
        \item core points: $\core = \{p\in \mathcal{I}: |\mathcal{N}_\epsilon(p)| \geq M \}$.
        \item border points: $\border = \{p \in \mathcal{I}-\core: \exists\, q \in \core\, s.t. \,p\in\Neps(q)\}$.
        \item noise points: $\noise = \mathcal{I} - \core - \border$.
    \end{itemize}
\end{definition}

DBSCAN uses three pairwise-point relations, which we give below in Definitions \ref{def:eps_direct}, \ref{def:eps_reach} and \ref{def:eps_connect}. 
\begin{definition}[direct $\epsilon\mh\text{reachable}$]
\label{def:eps_direct}
    A point $q$ is directly $\epsilon$-reachable from a point $p$ if $p \in \core$, and $q \in \Neps (p)$, which is denoted by $p\directRadireach q$.
\end{definition}
\begin{definition}[$\epsilon\mh\text{reachable}$]
\label{def:eps_reach}
    A point $q$ is $\epsilon$-reachable from a point $p$ if $p \in \core$ and $\exists \, p_1, p_2, ..., p_n \in \core$, s.t. $p\directRadireach p_1$, $p_1\directRadireach p_2$, ..., $p_n\directRadireach q$ ,which is denoted by $p \radireach q$.
\end{definition}
\begin{definition}[$\epsilon\mh\text{connected}$]
\label{def:eps_connect}
   A point $p$ is $\epsilon\mh\text{connected}$ with $q$ if $\exists \, o\in \core$, s.t. $o\radireach p$ and $o \radireach q$, which is denoted by $p \radicon q$.
\end{definition}

Figure \ref{fig:dbscan} illustrates the basic concepts of DBSCAN. Using these definitions, a density-based cluster in DBSCAN given by  Definition \ref{def:clusterdbscan}.
\begin{definition}[cluster of DBSCAN] \label{def:clusterdbscan}
A DBSCAN cluster $C$ is a nonempty subset of points in $\mathcal{I}$ satisfying that:
\begin{enumerate}[label=(\roman*)]
    \item  $\forall \, p, q \in \mathcal{I}$: if $p\in C$ and $p\radireach q$, then $q \in C$. (Maximality) 
    \item $\forall \, p,q \in C$: $p \radicon q$. (Connectivity)
\end{enumerate}
\end{definition}

For core points, the relation of $\epsilon\mh$reachable is equivalent to the relation of $\epsilon\mh$connected, and this relation is an \ipoint{equivalence relation}, i.e., it's symmetric, transitive and reflexive. By Definition \ref{def:clusterdbscan}, the DBSCAN clusters restricted to core points are equivalence classes defined by the relation of $\epsilon \mh$reachable. Thus, a core point belongs to a unique cluster. By the connectivity in Definition \ref{def:clusterdbscan} (ii), a border point could belong to any cluster that contains core point in the border point's $\epsilon\mh$radius neighborhood. A noise point belongs to no cluster. Therefore, from the implementation point of view, we are interested in clustering core points.

\begin{figure}[tbp]
    \centering
\begin {tikzpicture}[xscale=0.6, yscale=0.6, -latex ,auto ,node distance =2 cm and 2cm,
semithick ,
core/.style ={ circle ,top color =red!80 , bottom color = red ,
draw,black , text=blue , minimum size =.01 cm,inner sep=-2pt},
border/.style ={ circle ,top color =green!80 , bottom color = green ,
draw,black , text=blue , minimum size =.01 cm,inner sep=-2pt},
noise/.style ={ circle ,top color =blue!80 , bottom color = blue ,
draw,black , text=blue , minimum size =.01 cm,inner sep=-2pt}]
\node[core,label=right:{\large $p$}] at (0,0) {};
\node[core,label=left:{\large $q$}] at (1.7,0) {};
\node[core] at (-0.35,-0.7) {};
\node[core] at (0.1,-0.4) {};
\node[core] at (-0.1,-0.6) {};
\node[core,label=above:{\large $r$}] at (2.3,0.1) {};
\node[core] at (2.4,-0.2) {};
\node[core] at (2.3,-0.6) {};
\node[border] at (-1.3,1.414) {};
\node[noise] at (3,2.3) {};
\node[noise] at (2.8,2.4) {};
\draw[red,thick,dashed] (0,0) circle (2cm);
\draw[red,thick,dashed] (1.7,0) circle (2cm);
\draw[->,line width=1pt] (0,0) -- (0.85,1.77);
\draw[->,line width=1pt] (1.7,0) -- (0.85,-1.77);
\node[label={\large $\epsilon$}] at (0.4,0.6) {};
\node[label={\large $\epsilon$}] at (1.4,-1.6) {};
\end{tikzpicture}
\caption{Key definitions of DBSCAN: the hyperparameters are $\epsilon$ and $M=3$ in this example. Red dots are \textcolor{red}{core} points,  green dot is \textcolor{green}{border} point, blue dots are \textcolor{blue}{noise} points. $q$ is directly $\epsilon\mh$reachable from $p$ since $q\in \mathcal{N}_\epsilon(p)$;  $q$ is not directly $M\mh$reachable from $p$ since $p\notin \mathcal{N}_{M}(q)$ and $q\notin \mathcal{N}_M(p)$. $q$ and $r$ are both directly $\epsilon\mh$reachable and directly $M\mh$reachable from each other.}
    \label{fig:dbscan}
\end{figure} 

\subsection{\textit{k}NN-DBSCAN}\label{s:knn-dbscan}

In \dbsc{}, we classify points exactly the same as in DBSCAN. In Lemma \ref{lemma:points_knn} we use  the $M$-nearest neighbor  of each point in $\mathcal{I}$ to define core points. From this Lemma it follows that  a $k$-NNG with $k\geq M$ is sufficient to classify points into core, border, and noise points.
\begin{lemma}
\label{lemma:points_knn}
   \indent (i) A point $p$ is a core point, i.e., $p \in \core$, iff $\max\limits_{q\in\mathcal{N}_M(p)} d(p,q)\leq \epsilon$ . \\
    \indent (ii) A point $p$ is a border point, i.e., $p \in \border$, iff $p\notin \core$ and $\,\exists \, q \in \core \cap \mathcal{N}_M(p)$, s.t. $d(p, q) \leq \epsilon$.\\
    \indent (iii) A point $p$ is a noise point, i.e., $p \in \noise$, iff $p\notin \core$ and $p\notin \border$.
\end{lemma}

In DBSCAN, a cluster is formed by connecting core points with its $\epsilon\mh$radius neighbors. But an $M$-nearest neighbor graph may be not. 
This is because for two core points $p$ and $q$,  $p\in \Neps(q)$ does \ipoint{not} imply $p\in \mathcal{N}_{M}(q)$. An intuitive way of thinking this is that if we assume all points are core points, then the $M$-nearest neighborhood graph is a subgraph of $\epsilon$-radius neighborhood graph.

To enable $M\mh$nearest neighbor clustering in \dbsc{}, we modify the relation of direct $\epsilon \mh$reachable (Definition \ref{def:eps_direct}) to a new relation called direct $M\mh\text{reachable}$ in Definition \ref{def:k_direct}. Similar to $\epsilon\mh$reachable and $\epsilon\mh$connected in DBSCAN, we define corresponding relations in \dbsc{} as $M\mh\text{reachable}$ (Definitions \ref{def:k_reach}) and $M\mh$connected in (Definition \ref{def:k_connect}). 
\begin{definition}[direct $M\mh\text{reachable}$]\label{def:k_direct}
    A point $q$ is directly $M\mh\text{reachable}$ from $p$, denoted by $p\xrightarrow{M} q$, if $p$ and $q$ satisfy:
    \begin{enumerate}[label=(\roman*)]
        \item $p \in \core$.
        \item$q \in \mathcal{N}_M(p)$ \textbf{or} $p \in \mathcal{N}_M(q)$.
    \end{enumerate}
\end{definition}

\begin{definition}[$M\mh\text{reachable}$]
\label{def:k_reach}
    A point $q$ is $M\mh\text{reachable}$ from $p$ if $p\in \core$, and $\exists \, p_1, p_2, ..., p_n \in \core$, s.t. $p\xrightarrow{M} p_1$, $p_1\xrightarrow{M} p_2$, ..., $p_n\xrightarrow{M} q$, which is denoted by $p \xrightharpoonup{M} q$.
\end{definition}

\begin{definition}[$M\mh\text{connected}$]
\label{def:k_connect}
    A point $p$ is $M\mh$connected with $q$ if $\exists \, o\in \core$, s.t. $o\xrightharpoonup{M} p$ and $o \xrightharpoonup{M}q$, which is denoted by $p \xleftrightarrow{M} q$.
\end{definition}
If we restrict to core points, the $M\mh\text{reachable}$ relation is an equivalence relation and is the same as $M\mh\text{connected}$ relation. We conclude this property in the following lemma:
\begin{lemma}
\label{lemma:equivalence}
    For core points $\core$,\\
    \indent (i) $M\mh\text{reachable}$ is an equivalence relation.\\
    \indent (ii) $M\mh\text{reachable}$ is equivalent to $M\mh\text{connected}$.
\end{lemma}

What is the relationship between the pairwise-reachability definitions for $k$NN-DBSCAN and DBSCAN? It turns out that for core points, each relation in $k$NN-DBSCAN is stronger than the corresponding relation in DBSCAN in Lemma \ref{lemma:relations}. Figure \ref{fig:dbscan} provides an example to illustrate this.
\begin{lemma} 
\label{lemma:relations}
    Given the same input parameters of radius $\epsilon$ and minimum number of points $M$, $\forall p, q \in \core$, the following conditions hold:\\
    \indent (i) $p \xrightarrow{M} q$ $\Longrightarrow$ $p \directRadireach q$,\\
    \indent (ii) $p  \xrightharpoonup{M}q$ $\Longrightarrow$ $p \radireach q$,\\
    \indent (iii) $p \xleftrightarrow{M} q$ $\Longrightarrow$ $p \radicon q$.
\end{lemma}

Consequently, we can define a cluster in Definition \ref{def:knn_cluster} similar to DBSCAN, i.e. connecting points by defined relations. 

\begin{definition}[cluster of \textit{k}NN-DBSCAN]
\label{def:knn_cluster}
A $k$NN-DBSCAN cluster $C$ is a nonempty subset of points in $\mathcal{I}$ satisfying that:\\
\indent (i) $\forall \, p, q \in \mathcal{I}$: if $p\in C$ and $p \xrightharpoonup{M} q$, then $q \in C$. \\
\indent (ii) $\forall \, p,q \in C$: $p \xleftrightarrow{M} q$.
\end{definition}

The \dbsc{} cluster is well-defined in that the $M\mh\text{reachable}$ relation is an \textit{equivalence relation} w.r.t. core points. As for the relation between DBSCAN and \dbsc{}, we can deduce the general relationship between \dbsc{} and DBSCAN when the input are the same as follows.

\begin{theorem}[relations between \dbsc{} and DBSCAN]\label{thm:relation-to-dbscan} 
\indent Given the same input hyperparameters $\epsilon$ and $M$, we have:\\
\indent (i) A \dbsc{} cluster is a \textit{subset} of a DBSCAN cluster w.r.t. core points. Thus, the number of \dbsc{} clusters is \textbf{no less than} the number of DBSCAN clusters.\\\color{black}
\indent (ii) If the number of \dbsc{} clusters is the same as DBSCAN clusters, the clustering results of \dbsc{} \textbf{is identical to} the DBSCAN clustering results. 
\end{theorem}
\begin{proof}
    
\indent (i) It is sufficient to show that any two core points in a same \dbsc{} cluster belong to a same DBSCAN cluster. Let $p,\,q\in\core$ belong to a cluster in \dbsc{}. From \defref{def:knn_cluster}, $p\xleftrightarrow{M} q$. From \lemmaref{lemma:relations}, $p\xleftrightarrow{\epsilon} q$ and $p\xrightharpoonup{\epsilon} q$. Thus $p$ and $q$ lie in a same cluster in DBSCAN by condition (i) in \defref{def:clusterdbscan}.

\color{black}
\indent (ii) To prove the second claim, we can use a proof by contradiction. Assume that the clustering results of DBSCAN and \dbsc{} differ. From part (i) of the proof, we have already established that any two core points in the same \dbsc{} cluster also belong to the same DBSCAN cluster. Thus, there exits a DBSCAN cluster $C$ with core points $p,q\in C$, such that $p$ and $q$ belong to different \dbsc{} clusters, say $p \in C_1^\prime$ and $q \in C_2^\prime$. We will now demonstrate that this assumption leads to a contradiction.

Let $n$ be the number of DBSCAN cluster or \dbsc{} cluster on points $\mathcal{I}$. Define the remaining points after excluding those in $C_1^\prime $ and $C_2^\prime$ as $\mathcal{I}_{\text{rest}}$, i.e. $\mathcal{I}_{\text{rest}}=\mathcal{I} \backslash \{ C_1^\prime \cup C_2^\prime\} $. Now, consider applying \dbsc{} and DBSCAN to cluster the points in $\mathcal{I}_{\text{rest}}$ using the same input parameters. For \dbsc{}, it is straightforward to see that the number of clusters on $\mathcal{I}_{\text{rest}}$ is $n_{\text{rest}}^\prime = n-2$. For DBSCAN, since we already know from part (i) that $ C_1^\prime \cup C_2^\prime \subseteq C$, the number of DBSCAN clusters on $\mathcal{I}_{\text{rest}}$ can only be $n_{rest}\in\{n +1, n, n-1\}$, depending on the data. However, this implies that $n_{\text{rest}} > n_{\text{rest}}^\prime$. But this contradicts our earlier result in part (i), where we established that $n_{\text{rest}} \leq n_{\text{rest}}^\prime$. 
\end{proof}

\color{black}

Next, we discuss the relations between the MST of the \enn{} or \knn{} and the two DBSCAN algorithms.

\subsection{MST}\label{s:mst}
 Although MSTs have  been discussed in the context of DBSCAN, to our knowledge a direct relationship between these two has not been proven. 
The main result is that forming  MSTs and finding connected components are mathematically equivalent to the DBSCAN algorithm. Note however that the clusterings of DBSCAN and \dbsc{} are not the same.  For an undirected graph $G=(\mathcal{I},E)$ with edges assigned real-valued weights, a minimum spanning tree (MST) is a spanning acyclic subgraph of $G$ having the least total weight. If the input graph is not  connected, the MST comprises disconnected subtrees. A MST cluster is defined in \ref{def:mstcluster}. Since the subtrees of MST are disconnected, clusters of a MST form a clustering for $\MA{I}$.
\begin{definition}[MST cluster]
\label{def:mstcluster}
    Given a graph $G$, a MST cluster $C$ is an nonempty subset of $\mathcal{I}$ consisting all points of a subtree of the MST. 
\end{definition}

In the following, we use $G_\epsilon$ to denote the \enn{}, and $G_M$ to denote the $k$-NNG with $k=M$. We also use $G_{\epsilon,\text{core}}$ to denote the subgraph whose vertices are only the core points (as defined in DBSCAN); and $G_{M,\text{core}}$ to be the subgraph of core points (as defined in \dbsc{}). The theorem below is stated informally. 
\begin{theorem}
\label{thm:mst} 
\indent (i) For core points, clusters of MST w.r.t $G_{\epsilon,\text{core}}$ are the same as clusters of DBSCAN.\\
\indent (ii) For core points, clusters of MST w.r.t $G_{M,\text{core}}$ are the same as the clusters of $k$NN-DBSCAN.
\end{theorem}

\subsection{Inexact MST}\label{s:inexact-mst}

Constructing an MST not only yields the clustering results of DBSCAN or \dbsc{}, but can also be used for hierarchical clustering. Unfortunately, building such an exact MST in distributed memory settings can be very costly. Therefore, we introduce an approximate MST, referred to as an inexact MST. This MST provides the exact clustering results (\Cref{thm:inexactmst}) and can be efficiently constructed in a distributed memory setting (\secref{s:algorithms}).

Given a graph, we vertex-partition it into $l$ parts, i.e., $\mathcal{I} = \bigcup_{i=1}^{l} \mathcal{I}_i$. We denote the graph associated with points in group $i$ as $G_i = (\mathcal{I}_i, E_i)$. For each group $i$, edges $E_i$ are split into internal edges denoted by $E_{i,\text{local}}$, and edges connected with other groups denoted by $E_{i,\text{cut}}$, i.e., $E_i = E_{i,\text{local}}\cup E_{i,\text{cut}}$. 

We define a new spanning tree called inexact MST by the following steps. \ding{202} For each group $i$, we find $MST_{i,\text{local}}$ as the MST of subgraph $(\mathcal{I}_i, E_{i,\text{local}})$ which only contains the local edges of $i$. \ding{203} We treat each local subtree as a super vertex and denote $\widehat{\mathcal{I}}$ as the set of all such super vertices. Now $(\widehat{\mathcal{I}}, E_{\text{cut}} = \bigcup_{i=1}^l E_{i,\text{cut}})$ is a graph containing all the edges in the cut of the partitioning, and the MST obtained for such graph is denoted by $MST_{\text{cut}}$ or \idef{cut MST}. \ding{204} The inexact MST under such  partitioning setting is defined as the combination of all local MSTs and the cut MST, i.e., $MST_{\text{inexact}} = (\cup_{i=1}^l MST_{i,\text{local}}) \bigcup MST_{\text{cut}}$. An illustration of the inexact MST is shown in Figure \ref{fig:inexact_mst}.

\begin{figure}[tbp]
    \centering
\begin {tikzpicture}[xscale=0.7, yscale=0.6]
    \draw[thick,red,zigzag] (-2.5,1) coordinate (a1) -- (-1,1.5) coordinate (b1);
    \draw[thick,red,zigzag] (-2.9,-.2) coordinate (a1) -- (-0.6,-1.) coordinate (b1);
    \draw[thick,red,zigzag] (1,.35) coordinate (a1) -- (2.5,.35) coordinate (b1);
    \draw[thick,red,zigzag] (1,1.5) coordinate (a1) -- (2.8,1.5) coordinate (b1);
    \draw[thick,red,zigzag] (0.3,-1.) coordinate (a1) -- (2,-0.4) coordinate (b1);
    \draw[ultra thick,blue] (2,.31) coordinate (a1) -- (2,-0.4) coordinate (b1);
    \draw[ultra thick,blue] (-0.6,-1.) coordinate (a1) -- (0.3,-1) coordinate (b1);
    \draw (-1.7,0) ellipse (1.6cm and 2cm);
    \draw (2,1) ellipse (2cm and 0.9cm);
    \draw (2,-1) ellipse (2cm and 0.8cm);
    \draw[dashed] (-0.6,-1) circle (1cm);
    \draw[dashed] (2,-.4) circle (1cm);
    \draw[->,dashed,line width=1pt] (2,-.4) -- (3,-.4);
    \draw[->,dashed,line width=1pt] (-0.6,-1) -- (-0.8,-0.1);
    \node[label=$\epsilon$] at (-.5,-0.55) {};
    \node[label=$\epsilon$] at (2.5,-0.43) {};
    \node[] at (-2,0) {group 1};
    \node[] at (2.5,1) {group 2};
    \node[] at (3.,-1.2) {group 3};
\end{tikzpicture}
    \caption{An illustration of the inexact MST. We partition points into three groups and consider $\epsilon\mh$radius neighborhood graph  of these points, i.e., $G_\epsilon$.  The inexact MST is a spanning tree combined by the local MST and cut MST. In this example, the inexact MST has three disconnected subtrees.}
    \label{fig:inexact_mst}
\end{figure} 

Note that the inexact MST may not be the same as the exact MST since the cut edges are not considered when finding the local MSTs. But we can use the inexact MST to obtain the clusters of core points in DBSCAN or $k$NN-DBSCAN. We conclude this property in Theorem~\ref{thm:inexactmst}, which forms the basis of our parallel, distributed-memory \dbsc{} algorithm.  

\begin{theorem}\label{thm:inexactmst}
\indent (i) For core points, clusters of the inexact MST w.r.t $G_{\epsilon,\text{core}}$ and given partitioning setting are the same as clusters of DBSCAN.\\
\indent (ii) For core points, clusters of the inexact MST w.r.t $G_{M,\text{core}}$ and given partitioning setting are the same as the clusters of $k$NN-DBSCAN.
\end{theorem}

%% file: s2s_algos.tex
In this section, we present our hybrid MPI/OpenMP implementation of the $k$NN-DBSCAN algorithm. The algorithm takes as input the radius $\epsilon$, the minimum number of neighbors $M$, and the directed $k$-nearest neighbor graph $G_k$ with $k \geq M$. With $n$ points and $p$ processes in total, we evenly partition the points across each process. Each process stores a subgraph denoted as $\mathbf{G}$, containing $n/{p}$ points and a total of $k{n}/{p}$ edges. We use $\mathbf{G}[i,m]=\{i,\,j,\,w\}$ to represent the edge from point $i$ to its $m$-th neighbor point $j$ with weight $w$.

The  core idea of the $k$NN-DBSCAN follows from Theorem~\ref{thm:inexactmst}, which involves clustering the core points by constructing an inexact MST on the subgraph of core points. To obtain the inexact MST, we employ Boruvka's algorithm in both the local MST construction and the cut MST construction. The algorithm is divided into four stages: 1) Finding core points, 2) Local MST construction, 3) Cut MST construction, 4) Clustering border points.  A pseudocode of our parallel $k$NN-DBSCAN algorithm is presented in \Cref{alg:knnDBSCAN}. 

\textcolor{black}{We chose Boruvka's algorithm over Kruskal's algorithm~\cite{kruskal} because Boruvka's is easier to parallelize. In Kruskal's algorithm, all edges need to be sorted from cheapest to most expensive. At each step, the algorithm removes the minimum-weight edge and, if it does not create a cycle in the graph, adds it to the MST; otherwise, it discards the edge. In contrast, Boruvka's algorithm works by identifying the nearest neighbors for each component, repeatedly selecting the cheapest edge from each component, and adding it to the MST. Finding the cheapest outgoing edge from each component can be done efficiently in parallel. However, parallelizing Kruskal's algorithm is challenging because it requires maintaining and checking edges in a strict order, making explicit parallelism difficult to achieve.}

We provide the details of the local MST construction in \Cref{sec:local_mst} and the cut MST construction in \Cref{sec:cut_mst}. Additionally, we outline the complexity of our parallel implementation in \Cref{sec:complexity}.
\begin{algorithm}[h]
\small
    \caption{Parallel $k$NN-DBSCAN}\label{alg:knnDBSCAN}
\begin{spacing}{1.2}
\begin{algorithmic}[1]
    \Require radius $\epsilon$, minimum number of neighbors $M$, point indices $\mathcal{I}$, edges of $k$NN graph $\mathbf{G}$ ($k \geq M$)\
    \Ensure cluster labels of points $R$
    \LineComment{Select and mark core points}
    \State $\mathcal{I}_{\text{core}} \gets \emptyset$, $R[i]\gets -1,\,i \in \mathcal{I}$
    \ForEach{point $i\in \mathcal{I}$}{\textbf{ in parallel}}
        \If{$\mathbf{G}[i,M].w \leq \epsilon$}
            \State $\mathcal{I}_{\text{core}} \gets \mathcal{I}_{\text{core}}\cup \{ i\}$
            \State $R[i] \gets i$
        \EndIf
    \EndFor

     \LineComment{Construct local MST of core points}
    \State $R,\, E_{\text{cut}},\, \mathcal{T}$ = \textsc{ParallelLocalMST}($\mathbf{G},\,\mathcal{I}_{\text{core}},\,R$)
    
    \LineComment{Construct cut MST of core points}
    \State $R$ = \textsc{DistributedCutMST}($R,\, E_{\text{cut}},\, \mathcal{T}$)    
    \LineComment{Cluster border points}
    \State $R$ = \textsc{ClusterBorder}($R,\,\mathcal{I}, \,\mathcal{I}_{\text{core}}, \, \mathbf{G}$)
    \State \Return $R$
\end{algorithmic}
\end{spacing}
\end{algorithm}

\subsection{Finding core points}
 A point is a core point if the distance to its $k$th neighbor is no larger than $\epsilon$ (Line 4),
As the neighbors for each point are stored in ascending order by Euclidean distance, we can easily identify core points. 
We also mark a core point with its own index in the label list $R$ (Line 6).

\subsection{Local MST construction} \label{sec:local_mst}

\begin{algorithm}[!t]
\small
\caption{Parallel local MST} \label{alg:local_mst}
\begin{spacing}{1.2}
\begin{algorithmic}[1]
    \Require graph $\mathbf{G}$, core points $\mathcal{I}_{\text{core}}$, labels $R$
    \Ensure labels $R$, cut edges $E_{\text{cut}}$, subtrees $\mathcal{T}$
\Procedure{ParallelLocalMST}{$\mathbf{G}, \mathcal{I}_{\text{core}}, \,R$}
    \State $\mathcal{T}\gets \mathcal{I}_{\text{core}}$, $E_{\text{cut}}\gets \emptyset$
    \State $L[i]\gets 1$, $i \in \mathcal{I}_{\text{core}}$ \Comment{initialize neighbor (edge) NO. to check}
    \Do
\LineComment{Find minimum edges}
    \State $ E_{\text{min}}[u]\gets \{u, -1, \text{sentinel}\}$, $u \in \mathcal{T}$
    \ForEach{point $i\in \mathcal{I}_{\text{core}}$}{\textbf{ in parallel}}
        \Do
        \State $e \gets \mathbf{G}[i,L[i]]$
        \State $ifbreak$ = \textsc{Findmin}($e, E_{\text{min}}, E_{\text{cut}}, R, L[i]$)
        \doWhile{($L[i] \leq M$) $\wedge$ ($ifbreak$)}
    \EndFor
    
\LineComment{Break symmetry}
    \State $R_{\mathcal{T}}[u]\gets E_{\text{min}}[u].j,\, u\in \mathcal{T}$ \Comment{initialize subtrees' roots}
    \ForEach{subtree $u\in \mathcal{T}$}{\textbf{ in parallel}}
        \State $v \gets R_{\mathcal{T}}[u]$
    \If{($v = -1$) $\vee$ ($u = R_{\mathcal{T}}[v] \wedge u < v$)}
        \State $R_{\mathcal{T}}[u]\gets u$
    \EndIf
    \EndFor
\LineComment{Pointer jumping}
    \State $Flags[u] \gets 1$, $u\in \mathcal{T}$ 
    \Do
        \ForEach{subtree $u\in \mathcal{T}$}{\textbf{ in parallel}}
            \If{$R_{\mathcal{T}}[u] \neq R_{\mathcal{T}}[R_{\mathcal{T}}[u]] $}
                \State $R_{\mathcal{T}}[u]\gets R_{\mathcal{T}}[R_{\mathcal{T}}[u]]$
            \ElsIf{$Flags[u]>0$} 
                \State $Flags[u] = 0$ \Comment{root for $u$ is found}
            \EndIf
        \EndFor    
    \doWhile{$\sum_u Flags[u]$ is changed}

\LineComment{Break cycles if cycle exits}
    \If{$\sum_u Flags[u] > 0$}
        \State \textsc{breakCycles}($\mathcal{T}, \,Flags,\, E_{\text{min}},\, R_{\mathcal{T}}$)
    \EndIf 
\LineComment{Update labels $R$ and subtree list $\mathcal{T}$}
    \State $N_{\mathcal{T}}^{\text{old}}\gets | \mathcal{T}|$, $\mathcal{T} \gets \emptyset$
        \State  $R[i] \gets R_{\mathcal{T}}[R[i]]$, $\mathcal{T} \gets \mathcal{T}\cup \{R[i] \}$, $i\in\mathcal{I}_\text{core}$
    \doWhile{$|\mathcal{T}| < N_{\mathcal{T}}^{\text{old}}$}
    \State \Return $R,\, E_{\text{cut}},\, \mathcal{T}$
\EndProcedure
\end{algorithmic}
\end{spacing}
\end{algorithm}

\begin{algorithm}[!t]
\small
    \caption{Find minimum edges}\label{alg:findmin}
\begin{spacing}{1.2}
\begin{algorithmic}[1]
\Procedure{Findmin}{$e, E_{\text{min}}, E_{\text{cut}}, R, l$}
        \State $i \gets e.i$, $j \gets e.j$
        \If{$j$ lies in another process}
            \State $E_{\text{cut}}\gets E_{\text{cut}}\cup \{e\}$, $l\gets l+1$
        \ElsIf{$R[j] \neq -1$ and $R[j]\neq R[i]$}
            \State \textsc{pwrite}($E_{\text{min}}[R[i]], \, e$)
            \State \Return true
        \Else
            \State $l \gets l +1$
        \EndIf
        \State \Return false
\EndProcedure
\Procedure{pwrite}{$e_{\text{min}}, \, e$}
    \Do
        \State $w_{\text{old}}$ $\gets e_{\text{min}}.w$
    \doWhile{($e.w < w_{\text{old}}$) $\wedge$ (\textsc{LockFreeCAS}($e_{\text{min}}, w_{\text{old}}, e$))}
\EndProcedure
\Procedure{LockFreeCAS}{$e_{\text{min}}, w_{\text{old}}, e$}
    \If{$e_{\text{min}}.w = w_{\text{old}}$}
        \State $e_{\text{min}} \gets e$
        \State \Return true
    \Else
        \State \Return false
    \EndIf
\EndProcedure
\end{algorithmic}
\end{spacing}
\end{algorithm}

\begin{figure}[t]
    \centering
    \includegraphics[height=4cm,width=6cm]{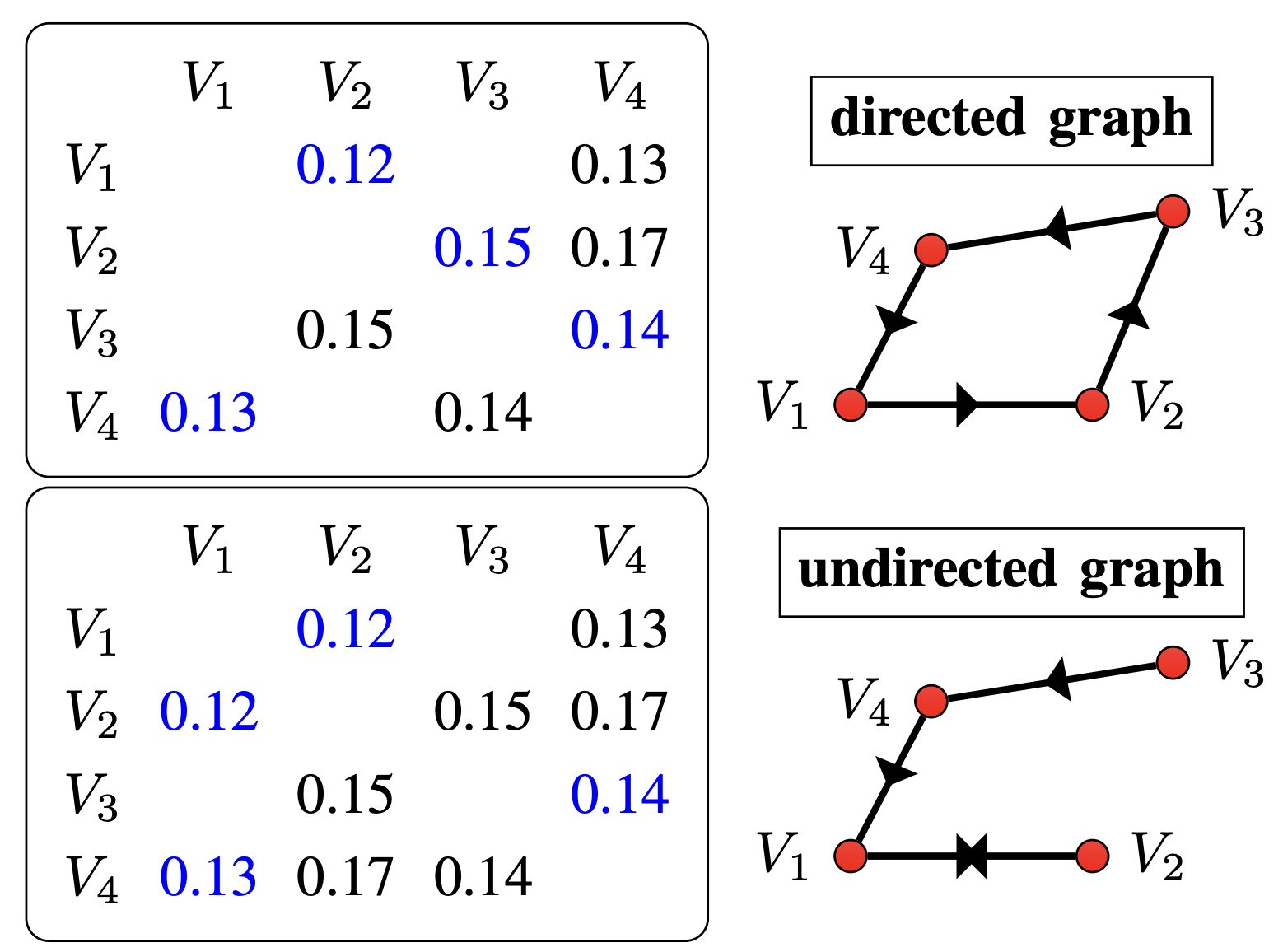}
    \caption{The figures demonstrate the formation of cycles in the directed approximate $k$NN graph. In the top-left table, we have a directed approximate $k$NN graph with 4 points and $k=2$. In the bottom-left table, the directed graph is converted into an undirected graph by symmetrizing the edges. The minimum edges are highlighted in blue in both tables. The bottom-right graph shows the minimum edges from the undirected graph, which will not contain cycles once we break symmetry between $V_1$ and $V_2$. On the other hand, as depicted in the top-right graph, a cycle may exist in the directed graph, and pointer jumping will never end in such a graph.}
    \label{fig:cycle}
\end{figure}

To construct the local MST, we implement Boruvka's method and  use OpenMP in each process since no communication is required in this step. The details of the algorithm are given in Algorithm \ref{alg:local_mst}. Boruvka's algorithm works by connecting different subtrees in an iterative way until none of them can be combined. At the beginning of the first iteration, subtree list $\mathcal{T}$ is initialized as $\mathcal{I}_\text{core}$. Each Boruvka's iteration has the following steps.\\
\indent\textit{Find minimum edges.}
First we need to find the outgoing edge with minimum weight for each subtree. To save work complexity, we avoid searching all edges in each iteration in our implementation. This is achieved by storing the number of neighbors searched in previous iterations in a list $L$ for each core point. For each candidate edge, a \textproc{FindMin} operation (see Algorithm \ref{alg:findmin}) is performed and three possibilities exist in this function. \ding{202} The edge may be a cut edge connecting a point owning by another process (Line 3). We append such edge to the cut edge list $E_\text{cut}$ for the later construction of cut MST. \ding{203} If the edge connects two core points that lie in \textit{different} subtrees (Line 5), it is a candidate for the minimum edge update. But a data race may occur when multiple threads update the minimum edge of a same subtree simultaneously. To address this problem, we use a function \textproc{pwrite} in Algorithm \ref{alg:findmin} with lock-free compare-and-swap operation (see function \textproc{LockFreeCAS} shown in Algorithm 
\ref{alg:findmin})~\cite{shun2013ligra}. The write loop stops when the intended new weight is not smaller than the weight stored in the minimum edge (Line 14). This approach guarantees that a minimum edge, say $E_\text{min}[u]$, does have the lightest weight over all outgoing \textit{local} edges of subtree $u\in\mathcal{T}$ after each iteration. \ding{204} If the edge points to a local non-core point or connects two core points belonging to a same subtree (Line 8), we need to check the next neighbor (if three is still any) of this point.\\
\indent\textit{Break symmetry.} 
Old subtrees connected by the selected minimum edges can be combined to form new larger subtrees. Our task then is to find a representative from each group of connected old subtrees as a \textit{root} for each new subtree. We first initialize a list $R_{\mathcal{T}}$ for old subtrees and store indices of points to which the minimum outgoing edges point (Line 13). If an old subtree has \textit{no} minimum outgoing edge, we claim itself as a root. If two old subtrees share the same minimum edge, we assign the one with smaller index as a root (Line 16-17).\\
\indent\textit{Pointer jumping and break cycles.} 
In this step, we update $R_{\mathcal{T}}$ for each old subtree to the root of the new subtree it belongs to. If an \textit{undirected} graph is used, no cycle exists once we break symmetries and this step can be finished by simple pointer jumping operations. We remark that using \textbf{\textit{directed approximate}} $k$NN graph, cycles may exist and pointer jumping operations may never end in cyclic graphs. This can be explained in the example shown in Figure \ref{fig:cycle}. To fix this issue, we need to detect if cycle exists and break it once it exits. The pointer jumping step stops only if there is \textit{no} new root found (Line 26). After this step, cycle(s) must exist if there are subtrees that have not found roots. These cycles are recovered and broken in a sequential way (Line 28-29). For the space consideration, we do not include the details of function \textproc{breakCycles}. In our experiments, small cycles are more likely to happen in low dimensional and dense data points at the beginning of the Boruvka's iterations. Breaking cycles does not affect the performance of our algorithm.\\
\indent\textit{Update labels and subtrees.} In the last step of each Boruvka's iteration, we update the root for every core point $i\in \mathcal{I}_{\text{core}}$ by using the root of the old subtree where $i$ belongs (Line 33). Furthermore, we update the list $\mathcal{T}$ to the roots of new subtrees. The Boruvka's iteration terminates when the total number of components is not changed (Line 34).

\begin{algorithm}[!t]
\small
    \caption{Distributed cut MST} \label{alg:cut_mst}
\begin{spacing}{1.2}
\begin{algorithmic}[1]
\Procedure{DistributedCutMST}{$E_{\text{cut}}, \,R,\, \mathcal{T}$}
\LineComment{Update cut edges}
        \ForEach{edge $e \in E_{\text{cut}}$}{\textbf{ in parallel}}
            \State $e.i \gets R[e.i]$
            \State update $e.j$ to label of $e.j$ \Comment{all to all exchange labels}
        \EndFor  
    \LineComment{Boruvka's iterations}
    \State $R_{\mathcal{T}}[u] \gets u,\, u\in \mathcal{T}$ 
    \State $\widehat{\mathcal{T}}\gets \mathcal{T}$, $\widehat{\mathcal{T}}_{l} \gets \mathcal{T}$, $N_{\text{root}}\gets$ MPI\_Allreduce($|\widehat{\mathcal{T}}_{l} |$)
    \Do
\LineComment{Find \textit{local} minimum cut edges $E_{\text{min}}^l$}
        \State $E_{\text{min}}^l[u]\gets \{u, -1, \text{sentinel}\}$, $u \in \widehat{\mathcal{T}}$
        \ForEach{edge $e \in E_{\text{cut}}$}{\textbf{ in parallel}}
            \If{$e.j \neq -1$} \Comment{only check active cut edge}
                \State \textsc{pwrite}($E_{\text{min}}^l[e.i], \, e$)
            \EndIf
        \EndFor          
\LineComment{Find \textit{global} minimum cut edges $E_{\text{min}}^g$}
        \State $E_{\text{recv}}\gets$ MPI\_Alltoallv($E_{\text{min}}^l$)
        \State $E_{\text{min}}^g[r]\gets \{r, -1, \text{sentinel}\}$, $r \in \widehat{\mathcal{T}}_{l}$
        \ForEach{edge $e \in E_{\text{recv}}$}{\textbf{ in parallel}}
                \State \textsc{pwrite}($E_{\text{min}}^g[e.i], \, e$)
        \EndFor          
\LineComment{Find roots of $\widehat{\mathcal{T}}_{l}$ and store in $R_l$}
        \State $R_l[r]\gets E_{\text{min}}^g[r].j$, $r \in \widehat{\mathcal{T}}_{l}$
        \State $R_l=$ \textsc{DistributedFindRoots}($R_l,\,\widehat{\mathcal{T}}_{l}$)
\LineComment{Update roots and cut edges}
        \State $\widehat{\mathcal{T}}\gets \emptyset$, $\widehat{\mathcal{T}}_{l}\gets \emptyset$, $N_{\text{root}}^\text{old}\gets N_{\text{root}}$
        \State $R_{\text{all}} \gets$ MPI\_Allgatherv($R_l$)
        \ForEach{subtree $u \in \mathcal{T}$}{\textbf{ in parallel}}
                \State $r \gets R_{\text{all}}[R_\mathcal{T}[u]]$, $R_\mathcal{T}[u] \gets r$, $\widehat{\mathcal{T}}\gets \widehat{\mathcal{T}}\cup \{r \}$
                \If{$r$ lies in local process}
                    \State $\widehat{\mathcal{T}}_{l} \gets \widehat{\mathcal{T}}_{l} \cup \{ r\}$
                \EndIf
        \EndFor 
        \State $N_{\text{root}}\gets$ MPI\_Allreduce($|\widehat{\mathcal{T}}_{l} |$)
        \ForEach{edge $e \in E_{\text{cut}}$}{\textbf{ in parallel}}
            \State $e.i \gets R_\text{all}[e.i]$, $e.j \gets R_\text{all}[e.j]$
            \If{$e.i = e.j$}
                \State $e.j = -1$ \Comment{edge becomes inactive}
            \EndIf
        \EndFor
    \doWhile{$N_\text{root} < N_{\text{root}}^\text{old} $ }
        \State $R[i] \gets R_{\mathcal{T}}[R[i]]$, $i\in \mathcal{I}_\text{core}$ \Comment{update core point's label}
    \State \Return $R$
\EndProcedure
\end{algorithmic}
\end{spacing}
\end{algorithm}

\begin{figure}[t]
    \centering
\begin{tikzpicture}[tree/.style={draw, minimum size=.5cm-\pgflinewidth, outer sep=0pt}]
    \node[] (a1) at (-2.5, 0.26) {$\widehat{\mathcal{T}}^0:$};
    \node[] (a2) at (-2.5, -.55) {$\widehat{\mathcal{T}}^1:$};
    \node[] (a3) at (-2.5, -1.45) {$\widehat{\mathcal{T}}_l^1:$};
     \draw[double,->] (a1)--(a2);
     \draw[double,->] (a2)--(a3);
    \node[tree, fill=t_green] at (-1.75,0.25) {\textcolor{red}{1}};
    \node[tree, fill=t_green] (t0) at (-1.25,0.25) {2};
    \node[tree, fill=t_yellow] at (-0.75,0.25) {\textcolor{red}{3}};
    \node[tree, fill=t_yellow] at (0.25,0.25) {4};
    \node[tree, fill=t_orange] (t1) at (.75,0.25) {\textcolor{red}{5}};
    \node[tree, fill=t_orange] at (1.25,0.25) {6};
    \node[tree, fill=t_blue] at (2.25,0.25) {\textcolor{red}{7}};
    \node[tree, fill=t_blue] (t2) at (2.75,0.25) {8};
    \node[tree, fill=t_red!70] at (3.25,0.25) {\textcolor{red}{9}};
    \node[tree, fill=t_red!70] at (4.25,0.25) {10};
    \node[tree, fill=t_green] (t3) at (4.75,0.25) {11};
    \node[tree, fill=t_green] at (5.25,0.25) {12};
    \node[tree, fill=t_green] at (-1.75,-.6) {1}; 
    \node[tree, fill=t_yellow] at (-1.25,-.6) {3};
    \node[tree, fill=t_yellow] at (0.25,-.6) {3};
    \node[tree, fill=t_orange] at (0.75,-.6) {5};
    \node[tree, fill=t_blue] at (2.25,-.6) {7};
    \node[tree, fill=t_red!70] at (2.75,-.6) {9};
    \node[tree, fill=t_red!70] at (4.25,-.6) {9};
    \node[tree, fill=t_green] at (4.75,-.6) {1};
    \node[tree, fill=t_green] at (-1.75,-1.45) {1}; 
    \node[tree, fill=t_yellow] at (-1.25,-1.45) {3};
    \node[tree, fill=t_orange] at (0.25,-1.45) {5};
    \node[tree, fill=t_blue] at (2.25,-1.45) {7};
    \node[tree, fill=t_red!70] at (2.75,-1.45) {9};
    \node[above=0.1cm of t0] {rank 0};
    \node[above=0.1cm of t1] {rank 1};
    \node[above=0.1cm of t2] {rank 2};
    \node[above=0.1cm of t3] {rank 3};
\end{tikzpicture}
    \caption{Illustration of distributed subtrees $\widehat{\mathcal{T}}$ and local roots $\widehat{\mathcal{T}}_l$ used in the first Boruvka's iteration of \textproc{DistributedCutMST} (see Algorithm \ref{alg:cut_mst}). We initialize 12 input local subtrees distributed in 4 MPI processes. After selecting minimum edges, subtrees are merged to form new ones. In the top row, old subtrees marked by the same color boxes belong to a new subtree now. We mark roots of newly formed subtrees with red numbers. In the last step of this iteration, $\widehat{\mathcal{T}}^0$ is updated to $\widehat{\mathcal{T}}^1$ by selecting all roots involved with each process. Root list $\widehat{\mathcal{T}}_l^1$ only consists of roots owned by each process.}
    \label{fig:cut_mst}
\end{figure}

\subsection{Cut MST construction}\label{sec:cut_mst}
The Boruvka's method is applied to construct cut MST for local subtrees $\mathcal{T}$ over all cut edges $E_{\text{cut}}$. From the previous local MST construction, we have known that for each cut edge $e\in E_\text{cut}$, $e.i$ and $e.j$ represent indices of the points connected by this edge. Thus, our first step is to change $e.i$ and $e.j$ into their labels. In particular, label of $e.i$ is easy to update by accessing local labels $R$, and label of $e.j$ is acquired by communicating with the owner process of $e.j$. Similar to the local MST construction, we combine old subtrees to form new subtrees by selecting minimum outgoing cut edges until they can no longer be merged. Thus, we initialize the subtree list $\widehat{\mathcal{T}}$ as the input local subtree list $\mathcal{T}$. But one difference in the distributed-memory case is that a subtree may have multiple (cut) edges belong to several processes. For the ease of implementation, we assign each subtree an owner process as the owner process of its representative. Thus, for each process we crate a new root list $\widehat{\mathcal{T}}_l$ to store elements from $\widehat{\mathcal{T}}$ whose owner is the local process. Figure \ref{fig:cut_mst} illustrates $\widehat{\mathcal{T}}$ and $\widehat{\mathcal{T}}_l$ in the first Boruvka's iteration of an example with 12 input subtrees in 4 MPI processes.

To handle the distributed nature of cut edges, the search for minimum edges is divided into two steps: \textit{local} and \textit{global}. In the local step, we identify local minimum edges from the cut edges within each subtree and store them in $E_{\text{min}}^l$. Next, we send the local minimum edges of each subtree to their respective owner processes, which can be achieved through an MPI all-to-all operation. The global minimum edge for each subtree $r\in \widehat{\mathcal{T}}_l$ is then determined by comparing and updating the recovered local minimums using the function \textproc{pwrite} (Line 19).

Similar to local MST construction, the next step consists of combining old subtrees using the selected minimum edges, finding representatives (roots) for the newly formed subtrees, and assigning roots to old subtrees. This can be done by breaking symmetry at first, then applying pointer jumping and breaking cycles if there is still any. The difference compared to the local case is that communications are required in the distributed-memory construction. Due to the space limitation, we do not include the implementation details of function \textproc{DistributedFindRoots} in the pseudocode. 

In the last step of each iteration, we update labels of local input subtrees, i.e. $R_{\mathcal{T}}$ to the new roots. Besides, the distributed subtree list $\widehat{\mathcal{T}}$ and local root list $\widehat{\mathcal{T}}_l$ are renewed for the next iteration. Figure \ref{fig:cut_mst} also illustrates how $\widehat{\mathcal{T}}$ and $\widehat{\mathcal{T}}_l$ are updated. We also inactivate cut edges that connect subtrees with the same roots (Line 34) to save work complexity in the following minimum edges' search. The termination condition of the Boruvka's iterations is the same as in the local MST construction, i.e., the total number of roots is no longer changed.


\subsection{Complexity analysis}\label{sec:complexity}
Since our parallel MST algorithm has two steps: local MST construction and cut MST construction, the total complexity of the algorithm is $ T_{\text{total}} =    T_{\text{local}} +   T_{\text{cut}}^{\text{work}} +   T_{\text{cut}}^{\text{commute}}$, where $T_{\text{local}}$ is the complexity of the local MST construction, $T_{\text{cut}}^{\text{compute}}$ and $T_{\text{cut}}^{\text{commute}}$ are work complexity and the communication complexity for cut MST construction. Let $n$ be the total number of points and $p$ be  the number of MPI processes. We consider $M$ neighbors for each point and assume that all points are core points. Thus, each process has $\tn\triangleq n/p$ points and $\tn M$ edges.

\textbf{Local MST construction}. In Algorithm \ref{alg:local_mst}, we assume a worst case scenario for the Boruvka's iteration: the total iteration number is $\log p$, and the number of subtrees ($|\mathcal{T}|$) at $i$th iteration is thus $\tn/{2^i}$. 

In the step of finding minimum edges, we need to check each edge at least once. Besides, the maximum possible number of edges used for compare and swap is $n$ at each iteration. Thus, the overall work complexity of this step is $\mathcal{O}({\tn M}+\tn\log \tn)$. 

To find roots of subtrees, breaking symmetry has overall complexity of $\sum_{i=1}^{\log \tn} \tn/{2^i} = \mathcal{O}(2\tn)$. In the pointer jumping step, the depth at $i$th iteration is less than $\log \frac{\tn}{2^i}$. Thus the overall complexity of pointer jumping is $\sum_{i=1}^{\log \tn}\frac{\tn}{2^i}\log \frac{\tn}{2^i} = \mathcal{O}(2\tn\log \tn)$. Updating root for each point requires $\mathcal{O}(\tn\log \tn)$ of work in total. 

The overall work complexity of local MST construction in each process is given by 
\begin{align*}
      T_{\text{local}} &= \mathcal{O}({\tn M}+\tn\log \tn) +  \mathcal{O}(2\tn \log \tn) + \mathcal{O}(\tn \log \tn) \nonumber\\
    &= \mathcal{O}({\tn M}+ 4\tn \log \tn) = \mathcal{O}\bigg({\frac{n}{p} M}+ 4\frac{n}{p} \log \frac{n}{p}\bigg).
\end{align*}

\textbf{Cut MST construction}. In order to discuss the complexity for cut MST construction, we assume that the number of subtrees formed in the local MST step is $\hat{n}$, and  the number of cut edges is $m_c$  in each process. Let $t_s$ be the communication latency, and $t_w$ be the reciprocal of the bandwidth. We use the hypercube topology for parallel computers to analyze the cut MST construction algorithm complexity.

At the beginning of Algorithm \ref{alg:cut_mst} (Line 3-5), updating cut edges requires $\mathcal{O}(2m_c)$ work and $\mathcal{O}((t_s+p m_c t_w){\log p})$ communication cost. For the Boruvka's iteration, we again assume the worst-case scenario and breakdown of the time into the following three components.

\textbf{Finding min-edges (Line 10-19 in \Cref{alg:cut_mst}).} The iteration number is $\log \hat{n}p$, and the number of components contained in each process at $i$th iteration is $\hat{n}/{2^i}$. Since the number of components reduced half after each iteration, we can assume that the number of active cut edges at $i$th iteration is $m_c/{2^i}$. Then sorting local minimum edges has overall complexity of $\mathcal{O}(2 m_c)$. In the step of finding global minimum edges, each iteration requires operation of \texttt{MPI\_Alltoall} with message size $\mathcal{O}(\hat{n}/{2^i})$. Thus the overall communication cost of this step is $\mathcal{O}(( t_s \log \hat{n}p +   t_w \hat{n}p)\log{p})$.

\textbf{Pointer jumping (Line 21-22 in \Cref{alg:cut_mst}).} The depth at each iteration can be assumed as $\log {p}$. Thus, the communication of this step has complexity of $\mathcal{O}( ( t_s \log \hat{n}p +   t_w \hat{n}p)\log{p})$. 

\textbf{Update $E_{\text{cut}}$ (Line 24-34 in \Cref{alg:cut_mst}).} To update cut edges, we need $\mathcal{O}(2m_c)$ of work and $\mathcal{O}(t_s \log p \log  \hat{n}p +2  t_w \hat{n}p)$ of communication cost for \texttt{MPI\_Allgather} and \texttt{MPI\_Allreduce}.

Based on the above complexity analysis, we can derive the work complexity of cut MST construction as
\begin{align*}
    T_{\text{cut}}^{\text{work}} = \mathcal{O}(6m_c),
\end{align*}
and the total communication cost as
\begin{align*}
    T_{\text{cut}}^{\text{commute}} = & \mathcal{O}\Big(({\log p} \log \hat{n}p)t_s + (m_c {p} + \hat{n}p\log {p})t_w \Big).
\end{align*}

%% file: s3_results.tex
\renewcommand{\arraystretch}{0.95}
\begin{table}[t]  
\color{black}
\small
\centering
\caption{\textcolor{black}{Datasets used for comparison between DBSCAN and \dbsc{} in~\Cref{s:acc}. The four 2D datasets on the left side of the table are visualized in \Cref{fig:2D}. The four datasets on the right side originally have higher dimensions, for which we applied dimension reduction techniques such as PCA or UMAP~\cite{mcinnes2018umap}.}}
\label{table:datasets-small} 
\setlength{\tabcolsep}{1pt} 
\begin{tabularx}{.8\columnwidth}{lCClCC}
\toprule
Dataset &$n$ &$d$ &Dataset &$n$ &$d$\\\midrule
Two Moons &800 &2 &KC House~\cite{kc_house} &21,613 &6\\
Aggregation~\cite{aggregation} &788 &2 &Credit Card~\cite{cc} &8,950 &6\\
Chameleon~\cite{chameleon}&8,000 &2 &Online Shoppers~\cite{online_shoppers} &12,330 &6\\
Worms~\cite{worms} &45,500 &2 &Human Activity~\cite{human_activity} &7,352 &5\\
\bottomrule
\end{tabularx}
\end{table}

\renewcommand{\arraystretch}{0.95}
\begin{table}[t]  
\small
\centering
\caption{Datasets used for comparison in \Cref{s:comparison}. $n$ and $d$ denote the total number and dimension of points in each data set. $\widetilde{r}_2$ and $\overline{r}_2$ denote the median and mean distances to the second nearest neighbor respectively. \textbf{MNIST70K} and \textbf{CIFAR-10} are real data sets with original full features. \textbf{MNIST2M} is a subset of the libsvm mnist8m nonlinearly projected to 50 dimensions using~\cite{mcinnes2018umap}. \textbf{Uniform1S} is generated randomly from a 20D sphere. \textbf{Uniform2S} comprises randomly distributed points  on two sphere surfaces: 1M points on a sphere of radius 0.1 and 3M points a sphere of radius 1.0.}
\label{table:datasets} 
\setlength{\tabcolsep}{1pt} 
\begin{tabularx}{.8\columnwidth}{l*{5}{C}}
\toprule
Dataset &$n$ &$d$ &$\widetilde{r}_2$ &$\overline{r}_2$ &\#clusters\\\midrule
\cellcolor{black!10}\textbf{MNIST70K} &70K &784 &1,082 &1,055 &10\\
\cellcolor{black!10}\textbf{MNIST2M} &2M  &50 &0.014 &0.018 &10\\
\cellcolor{black!10}\textbf{CIFAR-10} &50K  &3,072 &2,360 &2,350 &10\\
\cellcolor{black!10}\textbf{Uniform1S{ }} &4M  &20 &0.51 &0.51 &1\\
\cellcolor{black!10}\textbf{Uniform2S{ }} &4M  &10 &0.23 &0.22 &2\\
\bottomrule
\end{tabularx}
\end{table}

\begin{table}[!t]
    \centering
    \caption{Wall-clock time for computation of $k$-nearest neighbors by GOFMM on some of the synthetic data sets. $N$ is the number of points, $d$ represents dimension, $k_{\max}$ is the number of neighbors. The low dimensional sets with $d\leq 5$, GOFMM converges in about 10 iterations. The larger dimension run ($d=32$) requires 20 iterations.}
    \label{table:knngraph_time}
    \small
    \setlength{\tabcolsep}{1pt} 
    \begin{tabularx}{.7\columnwidth}{l*{4}{C}}
        \toprule
     $n(\times 10^6)$    & $d$ & $k_{\max}$ & \#cores &time (s)    \\ \midrule
    64   &3 &100 &1,792  & 363     \\ 
    64   &5 &100  &1,792   & 379   \\ 
    64  &32 &100  & 1,792  & 926    \\ 
    256   &3 &100 &7,168  & 583     \\ 
    256   &5 &100  &7,168   & 628   \\ \bottomrule
    \end{tabularx}
\end{table}

In our experiments, we attempt to answer the following questions regarding \dbsc{}.How does the clustering performance of \dbsc{} compare to that of DBSCAN? How well does our parallel implementation of \dbsc{} scale? How does its performance vary with the dimensionality of the dataset? In \Cref{s:acc}, we compare \dbsc{} with DBSCAN using small-scale datasets. In \Cref{s:scaling}, we present the strong and weak scaling results of our parallel implementation of \dbsc{} using synthetic datasets. In \Cref{s:comparison}, we compare the performance of \dbsc{} with a parallel implementation of DBSCAN.

\textbf{Datasets.} {For comparison between clustering results between DBSCAN and \dbsc{}, we use small scale datasets presented in \Cref{table:datasets-small}. We use synthetic data uniformly distributed across unit sphere surfaces in various dimensions to investigate the algorithm's scalability, as discussed in \Cref{s:scaling}. Furthermore, we use a combination of three real-world datasets and two synthetic datasets, as outlined in \Cref{table:datasets}, for the purpose of comparing \dbsc{} with an existing parallel DBSCAN implementation, detailed in \Cref{s:comparison}.}

\textbf{Implementation details and experimental setup.} All of our experiments were conducted on the Frontera system at the Texas Advanced Computing Center (TACC). Frontera CLX cluster has 8,008 dual-socket compute nodes, each with two 28 core 2.7GHz CPUs and 192GB memory. The parallel $k$NN-DBSCAN is written in \texttt{C++} and compiled with \texttt{intel-19 -O3}
. We use OpenMP for the shared memory parallelism and Intel MPI for distributed memory parallelism. Each MPI process loads a random $n/p$ portion of the dataset, {where $n$ is the total number of points,  $p$ is the number of MPI processes.}

{\textbf{Hyperparameters choice.}  The main parameters in our implementation are the number of nearest neighbors $k_{\max}$ in the \knn{} graph,  $\epsilon$,  and $M$ (the number of neighbors required to define a core point).  According to our definition of \dbsc{}, we can select $\epsilon$ and $M$ in a manner analogous to DBSCAN, as both methods share the same core and noisy point definitions. The determination of these parameters in DBSCAN involves employing a set of heuristics outlined in~\cite{ester1996dbscan,sander-1998density}, which rely on the statistics of nearest-neighbor distances and the intrinsic dimension of the dataset. Notably, the more straightforward parameter to set in DBSCAN is the minPts parameter $M$. Subsequently, $\epsilon$ can be determined by sorted $M$-distance plots~\cite{schubert-2017dbscan}. We remark that \dbsc{} can accelerate the computations regarding the choice of these parameters since varying $\epsilon$ doesn't require repeated nearest-neighbor range searches. The details are discussed in~\Cref{s:comparison}. In our synthetic experiments in~\Cref{s:scaling}, we opt for hyperparameter choices ensuring that all points are core points, primarily for scalability testing purposes. In the  experiments discussed in~\Cref{s:comparison}, we explore diverse sets of hyperparameters. }

\color{black}
\subsection{Clustering Performance: \dbsc{} vs. DBSCAN}\label{s:acc}

\begin{figure}[!t]
    \centering
\begin{tikzpicture}
    \node[draw, inner sep=0pt] (legend) at (6,2) {\includegraphics[width=5cm]{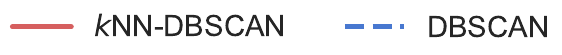}};
    \node[draw,rounded corners,fill=black!10]  at (2,1.45) {\small KC House};
    \node[inner sep=0pt] (a1) at (0,0) {\includegraphics[width=3.5cm]{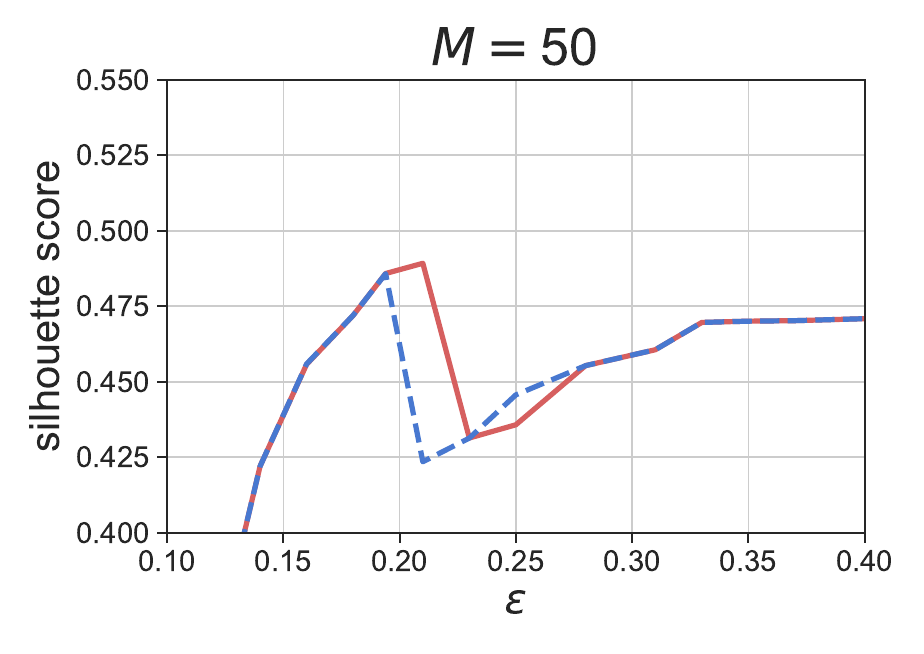}};
    \node[inner sep=0pt] (a2) at (3.5,0) {\includegraphics[width=3.5cm]{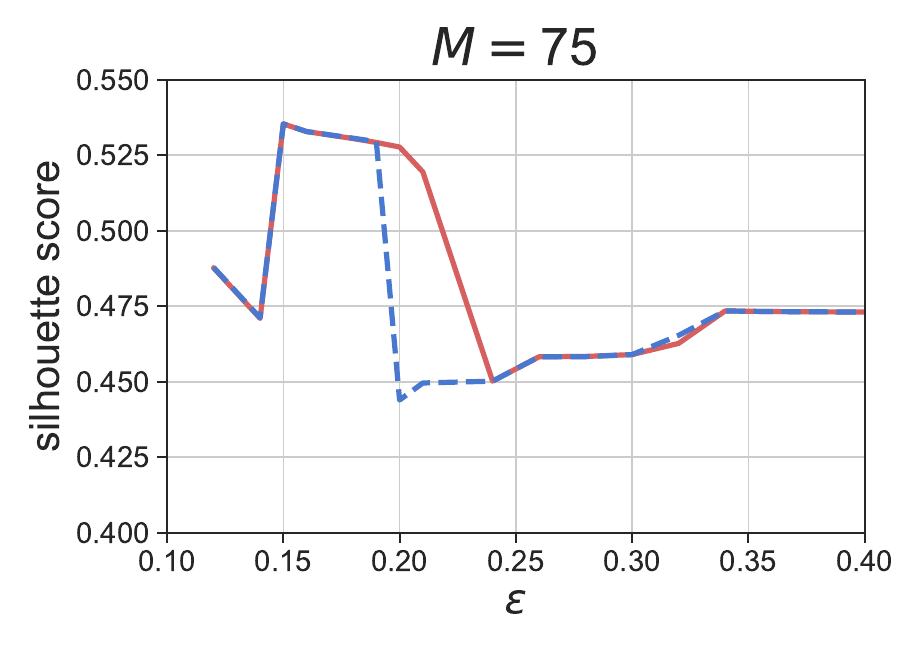}};
    \node[draw,rounded corners,fill=black!10]  at (9.5,1.45) {\small Credit Card};
    \node[inner sep=0pt] (b1) at (7.5,0) {\includegraphics[width=3.5cm]{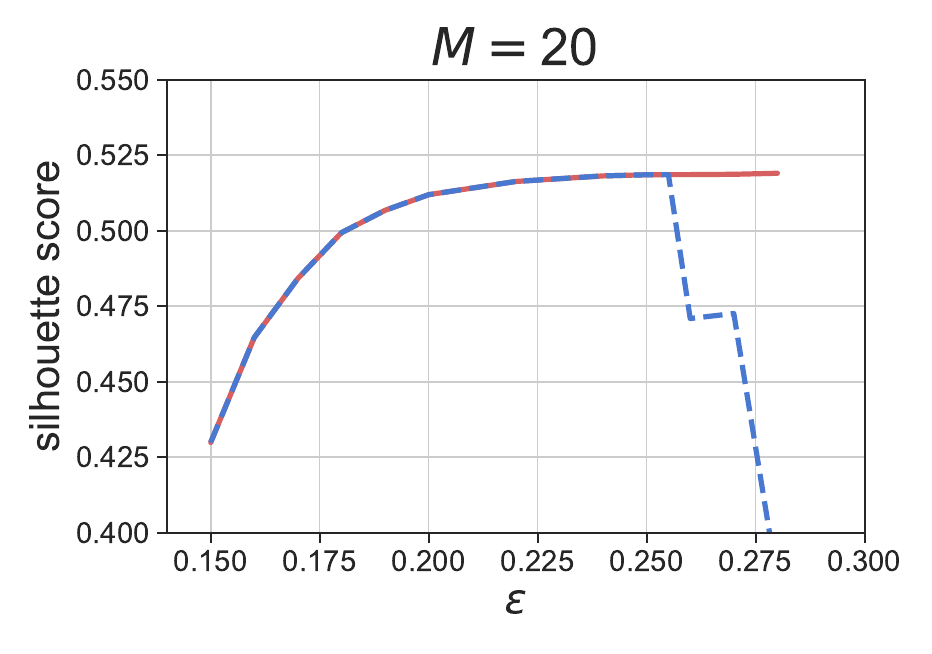}};
    \node[inner sep=0pt] (b2) at (11,0) {\includegraphics[width=3.5cm]{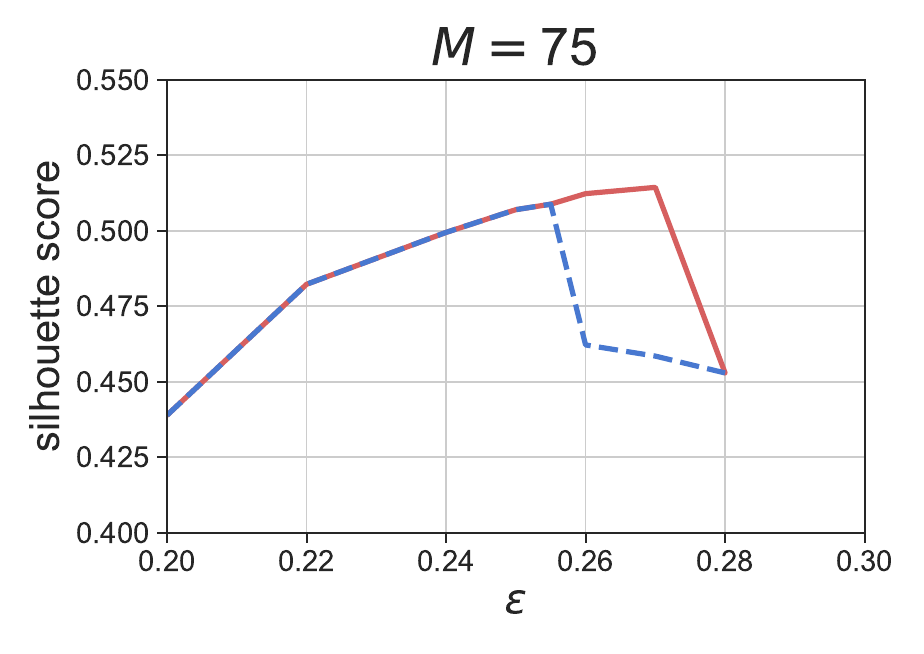}};
    \node[draw,rounded corners,fill=black!10]  at (2,-1.5) {\small Online Shoppers};
    \node[inner sep=0pt] (c1) at (0,-3) {\includegraphics[width=3.5cm]{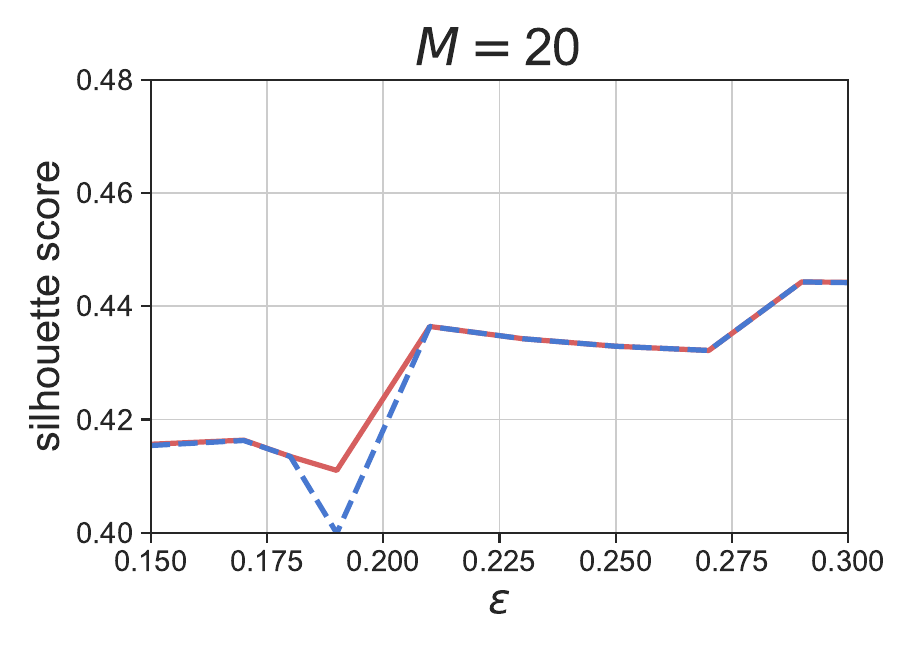}};
    \node[inner sep=0pt] (c2) at (3.5,-3) {\includegraphics[width=3.5cm]{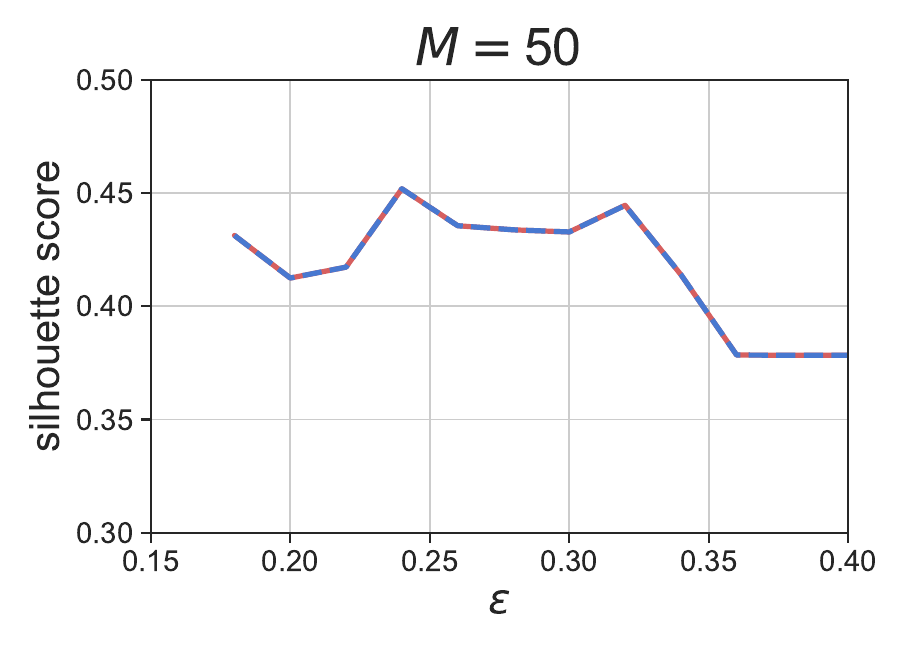}};
    \node[inner sep=0pt] (b1) at (7.5,-3) {\includegraphics[width=3.5cm]{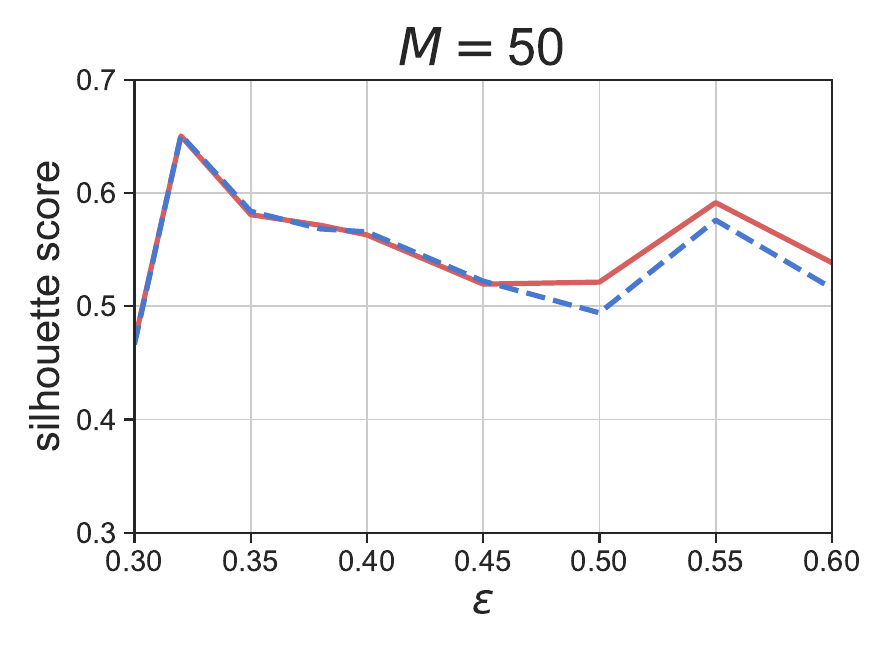}};
    \node[inner sep=0pt] (b2) at (11,-3) {\includegraphics[width=3.5cm]{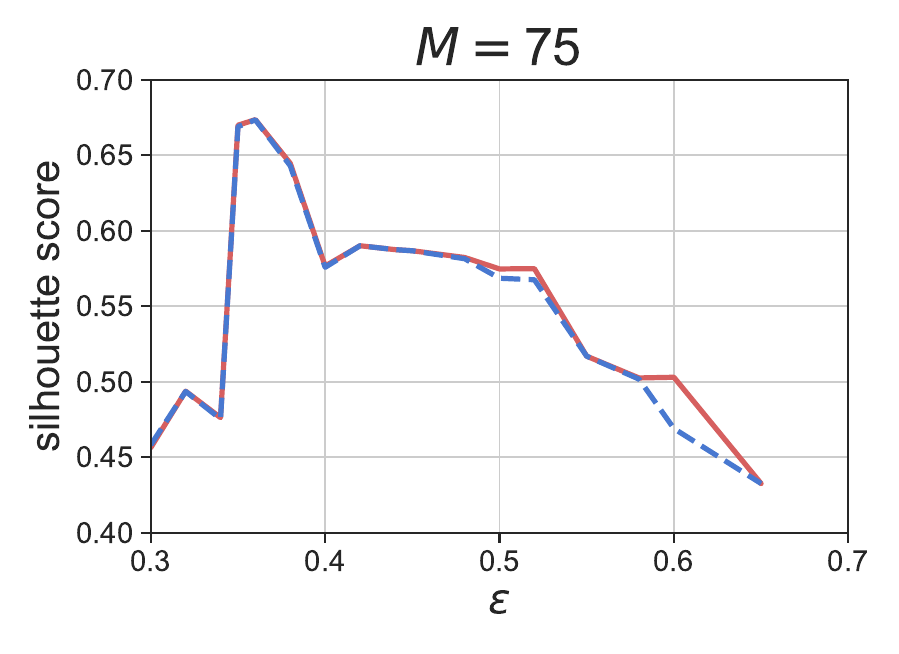}};
    \node[draw,rounded corners,fill=black!10]  at (9.5,-1.5) {\small Human Activity};
\draw[black!40,dashed] (5.5,1.5) -- (5.5,-4);
\end{tikzpicture}
    \caption{\textcolor{black}{Silhouette scores for the clustering results of DBSCAN and \dbsc{} on four small-scale datasets. Details for each dataset are provided in~\Cref{table:datasets-small}.}}
    \label{fig:cluster_ss}
\end{figure}

We present the clustering results on 2D datasets in \Cref{fig:2D}. It can be seen that both DBSCAN and \dbsc{} produce identical results when given the same input values. Additionally, it's evident that DBSCAN or \dbsc{} outperforms K-Means under these datasets. 

\Cref{fig:cluster_ss} displays the silhouette scores of DBSCAN and \dbsc{} across four datasets under different hyperparameters. The definition of the silhouette score is provided in \Cref{sec
}. The best possible score is 1, while the worst is -1. Scores near 0 indicate overlapping clusters, and negative values typically suggest that a sample has been incorrectly assigned to a cluster, as it is more similar to a different cluster. It can be observed that DBSCAN and \dbsc{} produce clusters with nearly identical silhouette scores on the Online Shoppers and Human Activity datasets. However, \dbsc{} outperforms DBSCAN when $\epsilon \in (0.2, 0.25)$ on the KC House dataset and when $\epsilon > 0.25$ on the Credit Card dataset.

\color{black}


\subsection{Scaling Results} \label{s:scaling}

First, to give an impression of the time required to construct the \knn{} with GOFMM, we report representative numbers in~\Cref{table:knngraph_time}. In all cases we try to compute the neighbors up to 99\% accuracy (estimated by sampling an $\bigO(1)$ subset from $\MA{I}$ and comparing the GOFMM solution with exhaustive search). While GOFMM has several parameters, which we did not try to optimize.  The key take-away from \Cref{table:knngraph_time}  is that compared to the time spent in \dbsc{}, constructing \knn{} is the dominant cost. However, in a practical setting  we need to do several \dbsc{} runs for hyperparameter search ($\epsilon$ and $M$ in \dbsc{});  so the \knn{} is amortized for any $M$ less than the nearest neighbor number considered in constructing \knn{}. In our \dbsc{} runs, we only report the timing for the parameters that yield the best clustering quality.

\begin{figure}[tb!]
    \centering
\begin{tikzpicture}
    \node[inner sep=0pt] (a) at (1.8,0) {\includegraphics[width=3.75cm]{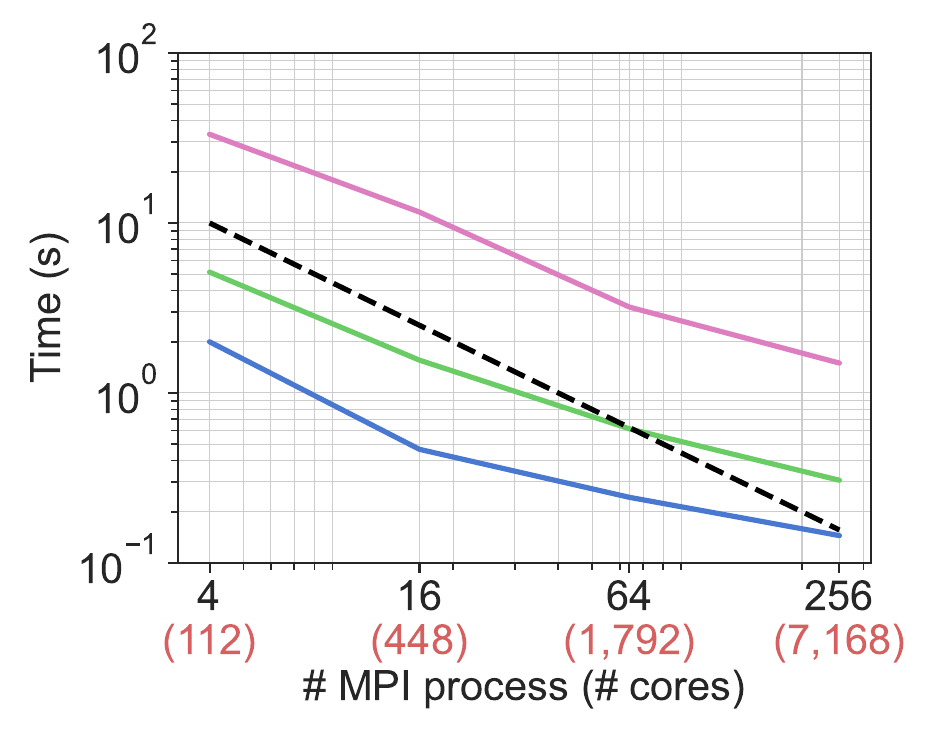}};
    \node[inner sep=0pt] (b) at (5.4,0) {\includegraphics[width=3.75cm]{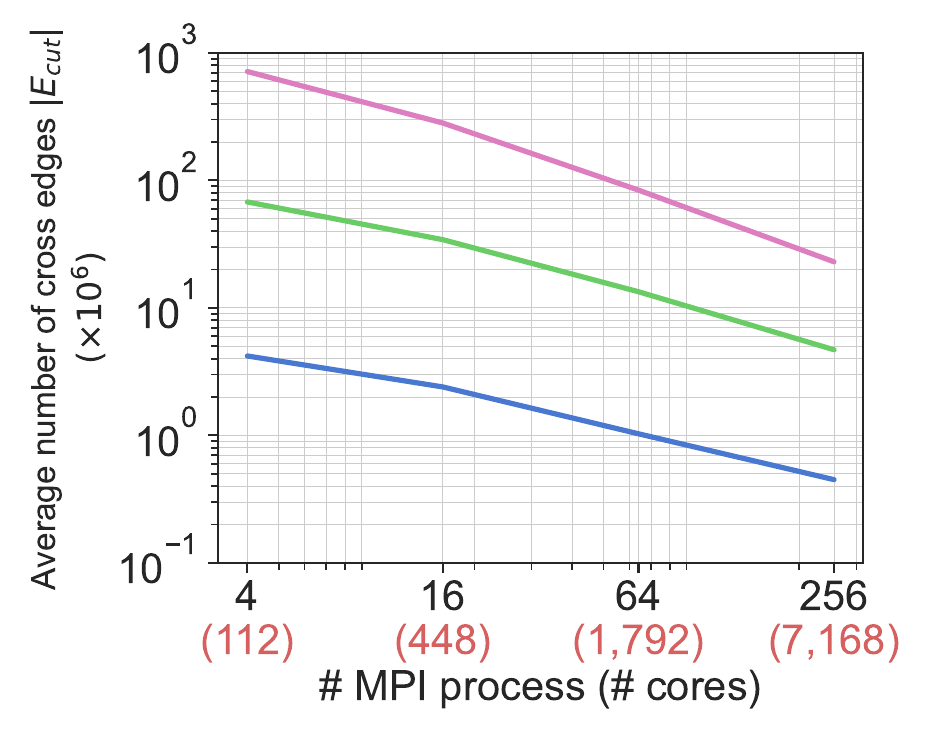}};
    \node[inner sep=0pt] (c) at (9,0) {\includegraphics[width=2cm]{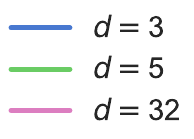}};
    \node[inner sep=0pt] (b) at (4,-3.5) {\includegraphics[width=15cm]{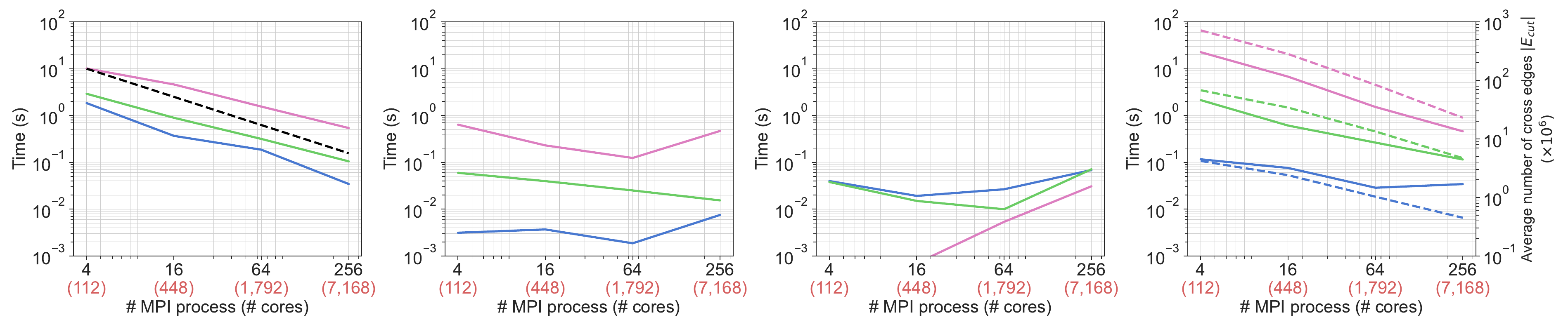}};
     \node[draw,rounded corners,fill=blue!20] at (-1.3,-1.8) {\small local};
     \node[draw,rounded corners,fill=blue!20]    at (2,-1.8) {\small min-edges};
     \node[draw,rounded corners,fill=blue!20]   at (5.6,-1.8) {\small pointer jumping};
     \node[draw,rounded corners,fill=blue!20]  at (9.2,-1.8) {\small update $E_{\text{cut}}$};
     \node[draw,rounded corners,fill=blue!20]  at (2.2,1.7) {\small total time};
     \node[draw,rounded corners,fill=blue!20]  at (5.6,1.7) {\small \# cut edges/MPI};
\end{tikzpicture}
    \caption{\color{black}Strong scaling results on 3D, 5D, and 32D datasets with 64 million points, ranging from 4 to 256 MPI processes. The upper row of plots shows the total wall-clock time and the average number of cut edges $|E_{\text{cut}}|$ per process. The lower row presents the breakdown of the wall-clock time. The red numbers along the x-axis indicate the number of cores used in each test. Note that the reported time in the plots corresponds to the MST construction time, given the $k$-NNG. The black dashed lines in the ``total time'' and ``local'' plots represent the ideal scaling trend. The dashed lines in the ``update $E_{\text{cut}}$'' plot indicate the average number of cut edges $|E_{\text{cut}}|$ per process.
    \color{black}}
    \label{fig:strong_scale}
\end{figure}

\begin{figure*}[tb!]
 \center
\begin{minipage}{.33\linewidth}
\begin{tikzpicture}[xscale = 0.6,yscale = 0.6]
\pgfplotsset{
    every tick label/.append style={font=\footnotesize},
    symbolic x coords={16M,64M,256M,512M,{1,024M}},
    xtick=data,
    x tick label style={rotate=45,anchor=east},
}
\begin{axis}[
    axis y line*=left,
    ybar stacked,
    bar width=15pt,
    nodes near coords,
    enlarge x limits=0.15,
    ymin=0., ymax= 0.8,
    point meta=explicit symbolic,
    legend style={at={(0.5,1.2)},
    anchor=north,legend columns=-1},
    ytick={0,0.2,...,0.8},
    ylabel={Time (seconds)},
    ticklabel style={font=\large},
    ]
\addplot+[ybar] table[x=x,y=t1] {plots/weak_3D.txt};
\addplot+[ybar] table[x=x,y=t2] {plots/weak_3D.txt};
\addplot+[ybar] table[x=x,y=t3] {plots/weak_3D.txt};
\addplot+[ybar] table[x=x,y=t4] {plots/weak_3D.txt};
\legend{\strut local, \strut min-edges, \strut pointer jumping, \strut update $E_{\text{cut}}$}
\end{axis}
\begin{axis}[
    axis y line*=right,
    axis x line=none,
    ymin = 0, ymax = 1.5,
    ytick = {0,.3,...,1.5},
    legend style={at={(0.8,.9)},
    draw = none,
    anchor=east,legend columns=1},
    ticklabel style={font=\large},
    ]
\addplot[blue!60,mark=o,mark size=3,line width=2pt] table[x = x,y=E_avg] {plots/weak_3D.txt};
\legend{\strut $|E_{\text{cut}}|$}
\end{axis}
\node[rotate=45,text=red!60] at (.9,-.5) {\scriptsize(448)};
\node[rotate=45,text=red!60] at (2.3,-.5) {\scriptsize(1,792)};
\node[rotate=45,text=red!60] at (3.6,-.5) {\scriptsize(7,168)};
\node[rotate=45,text=red!60] at (4.9,-.5) {\scriptsize(14,336)};
\node[rotate=45,text=red!60] at (6.15,-.6) {\scriptsize(28,672)};
\end{tikzpicture}
    \subcaption{3 dimensions}
    \label{fig:weak_3D}
\end{minipage}
\begin{minipage}{.31\linewidth}
\begin{tikzpicture}[xscale = 0.6,yscale = 0.6]
\pgfplotsset{
    symbolic x coords={16M,64M,256M,512M,{1,024M}},
    every tick label/.append style={font=\footnotesize},
    xtick=data,
    x tick label style={rotate=45,anchor=east},
}
\begin{axis}[
    axis y line*=left,
    ybar stacked,
    bar width=15pt,
    nodes near coords,
    enlarge x limits=0.15,
    ymin=0.,ymax=1.5,
    point meta=explicit symbolic,
    legend style={at={(0.5,1.2)},
    anchor=north,legend columns=-1},
    ytick={0,0.3,...,1.5},
    ticklabel style={font=\large},
    ]
\addplot+[ybar] table[x=x,y=t1] {plots/weak_5D.txt};
\addplot+[ybar] table[x=x,y=t2] {plots/weak_5D.txt};
\addplot+[ybar] table[x=x,y=t3] {plots/weak_5D.txt};
\addplot+[ybar] table[x=x,y=t4] {plots/weak_5D.txt};
\legend{\strut local, \strut min-edges, \strut pointer jumping, \strut update $E_{\text{cut}}$}
\end{axis}
\begin{axis}[
    axis y line*=right,
    axis x line=none,
    ymin = 0, ymax = 20,
    ytick = {0,5,...,20},
    legend style={at={(0.8,.9)},
    draw = none,
    anchor=east,legend columns=1},
    ticklabel style={font=\large},
    ]
\addplot[blue!60,mark=o,mark size=3,line width=2pt] table[x = x,y=E_avg] {plots/weak_5D.txt};
\legend{\strut $|E_{\text{cut}}|$}
\end{axis}
\node[rotate=45,text=red!60] at (.9,-.5) {\scriptsize(448)};
\node[rotate=45,text=red!60] at (2.3,-.5) {\scriptsize(1,792)};
\node[rotate=45,text=red!60] at (3.6,-.5) {\scriptsize(7,168)};
\node[rotate=45,text=red!60] at (4.9,-.5) {\scriptsize(14,336)};
\node[rotate=50,text=red!60] at (6.15,-.6) {\scriptsize(28,672)};
\end{tikzpicture}
    \subcaption{5 dimensions}
    \label{fig:weak_5D}
\end{minipage}
\begin{minipage}{.3\linewidth}
\begin{tikzpicture}[xscale = 0.6,yscale=0.6]
\pgfplotsset{
    every tick label/.append style={font=\footnotesize},
    every tick label/.append style={font=\footnotesize},
    symbolic x coords={128M, 512M, {2,048M}, {8,192M}, {65,536M}},
    xtick=data, x tick label style={rotate=45,anchor=east},
}
\begin{axis}[
    axis y line*=left,
    ybar stacked,
    bar width=15pt,
    nodes near coords,
    enlarge x limits=0.15,
    ymin=0.,ymax=50,
    point meta=explicit symbolic,
    legend style={at={(0.5,1.2)},
    anchor=north,legend columns=-1},
    ytick={0,10,...,50},
   ticklabel style={font=\large},
    ]
\addplot+[ybar] table[x=x,y=t1] {plots/weak_20D.txt};
\addplot+[ybar] table[x=x,y=t2] {plots/weak_20D.txt};
\addplot+[ybar] table[x=x,y=t3] {plots/weak_20D.txt};
\addplot+[ybar] table[x=x,y=t4] {plots/weak_20D.txt};
\legend{\strut local, \strut min-edges, \strut pointer jumping, \strut update $E_{\text{cut}}$}
\end{axis}
\begin{axis}[
    axis y line*=right,
    axis x line=none,
    ymin = 0, ymax = 1000,
    ytick = {0,200,...,1000},
    legend style={at={(0.55,.9)},
    draw = none,
    anchor=east,legend columns=1},
    ylabel ={Average number of edges in the graph cut ($\times10^6$)},
    ticklabel style={font=\large},
    ]
\addplot[blue!60,mark=o,mark size=3,line width=2pt] table[x = x,y=E_avg] {plots/weak_20D.txt};
\legend{\strut $|E_{\text{cut}}|$}
\end{axis}
\node[rotate=47,text=red!60] at (.9,-.5) {\scriptsize(224)};
\node[rotate=45,text=red!60] at (2.3,-.5) {\scriptsize(896)};
\node[rotate=45,text=red!60] at (3.6,-.5) {\scriptsize(3,584)};
\node[rotate=45,text=red!60] at (4.9,-.5) {\scriptsize(14,336)};
\node[rotate=50,text=red!60] at (6.15,-.6) {\scriptsize(114,688)};
\end{tikzpicture}
    \subcaption{20 dimensions}
    \label{fig:weak_20D}
\end{minipage}
  \caption{Weak scaling results on synthetic datasets. Each  test is conducted with 28 threads/MPI process, 2 MPI processes/node. The numbers in red along the x-axis indicate the core count used for each test. For each plot, the left y-axis shows the breakdown of wall-clock time in seconds, and the right y-axis displays the average number of cut edges. \textbf{(a)} \textbf{3-dimensional} data ranging from 16 million to 1,024 million points (1 million points per MPI process). \textbf{(b)} \textbf{5-dimensional} data ranging from 16 million to 1,024 million points (1 million points per MPI process). \textbf{(c)} \textbf{20-dimensional} data ranging from 128 million to 65,536 million points (16 million points per MPI process). Note that the time reported in the plots represents the MST construction time given \knn{}. }
\end{figure*}


We discuss the weak and strong scaling results for our parallel $k$NN-DBSCAN algorithm. We report the average number of edges in the graph cut across all MPI ranks, denoted as $|E_{\text{cut}}|$. The total time for each test is broken down into the following components:
\begin{itemize}
\small
  \setlength\itemsep{0em}
    \item \idef{local}:  construction of local MST.
    \item \idef{min-edges}: searching minimum edges from edges in the graph cut in the construction of cut MST.
    \item \idef{pointer jumping}: finding the roots of each subtree (from the previous iteration) in the construction of cut MST.
    \item \idef{update $E_{\text{cut}}$}: communication for update edges in the graph cut at the very beginning of cut MST construction and at the end of each Boruvka's iteration.
\end{itemize}


\textbf{Strong scaling.}  We present the strong scaling results in \Cref{fig:strong_scale} from experiments on 3-, 5-, and 32-dimensional datasets with 64 million points, using 4 to 256 MPI processes. Using the tests with 4 MPI processes as baselines, the speedups achieved with 256 MPI processes are 13.8x for the 3-dimensional dataset, 16.8x for the 5-dimensional dataset, and 22.2x for the 32-dimensional dataset. 

\color{black}
The ``total time'' plot shows that both the 5D and 32D results scale closely to the ideal trend. For the 3D test, the speedup is optimal when the number of MPI processes is low (e.g., from 4 to 16), but falls short of the ideal case when using 256 MPI processes. This can be explained from the lower row of plots in \Cref{fig:strong_scale}: with 256 MPI processes, the problem size becomes very small, causing communication times, such as pointer jumping, to dominate. One effect of increasing the dataset dimension on our implementation's performance is the rise in the number of cut edges across different MPI ranks, as shown in the ``\# cut edges/MPI'' plot.

The breakdown of total time, as shown in the plots, aligns with our complexity analysis in \Cref{sec:complexity}. We observe that the local MST time (in ``local'' plot) scales closely to the ideal trend.  According to our complexity analysis of cut MST, the communication time for finding minimum edges primarily depends on the number of cut edges per process and the number of Boruvka's iterations, and pointer jumping time depends on number of local MSTs and the number of Boruvka's iterations. As the number of MPI processes increases, the number of Boruvka's iterations rises, but the number of local MSTs and cut edges decreases. This explains why in some tests, the time for ``in-edges'' and ``pointer jumping'' initially decreases and then increases as the number of MPI processes grows. For updating $E_{\text{cut}}$, our analysis shows that its work complexity scales with $|E_{\text{cut}}|$, which is consistent with the experimental results shown in the ``update $E_{\text{cut}}$'' plot.
\color{black}

\textbf{Weak scaling.} Our weak scaling results are presented in \Cref{fig:weak_3D,fig:weak_5D,fig:weak_20D} for 3-, 5-, and 32-dimensional datasets respectively. For the 3-dimension and 5-dimension experiments, we use a problem size of 1 million points per MPI process, 2 million points per node. The 20-dimension weak-scaling tests are conducted on a problem size of 16 million points per MPI process from 128 million points up to 65,536 million points. Note that in the 20 dimensional tests, we used the  \qq{128M}  data sets as a base, and duplicated it in other tests. For example, the  \qq{512M}  test uses 4 of the \qq{128M} data sets. Thus the averaged numbers of edges in the graph cut are all the same in \Cref{fig:weak_20D}.  This is an artificial test to check the weak scaling of the algorithm without changing the characteristics of the underlying dataset.

\color{black}
The weak scaling results are consistent with our complexity analysis in \Cref{sec:complexity}. We observe that the local MST time (blue bars) remains nearly constant as the number of points and MPI processes increases, which is expected since the local problem size stays the same regardless of the number of MPI processes. Additionally, the local number of MSTs and cut edges per MPI process remains relatively stable as MPI processes increase. However, the number of Boruvka's iterations increases, which explains the rise in pointer jumping time (tan bars) and min-edge time (red bars) as the number of MPI processes grows. For updating $E_{\text{cut}}$ (gray bars), the work complexity scales with the number of cut edges per process, but the communication time increases as the number of MPI processes increases. This aligns with our observations from the plots: as the dimension increases, the time for updating $E_{\text{cut}}$ becomes more dominant since $|E_{\text{cut}}|$ grows with higher dimensions. Specifically, in the \qq{1,024M} tests, the time spent updating edges in the graph cut accounts for 29.5\% of the total time in the 5-dimensional case, compared to just 13.7\% in the 3-dimensional test. 
\color{black}

%
%

\subsection{Comparison with PDBSCAN}\label{s:comparison}

\begin{table}[t!]  
\footnotesize
\centering
\caption{Comparison between PDBSCAN and $k$NN-DBSCAN in wall-clock time ($T$ in seconds) and clustering accuracy (NMI, higher is better) on \textbf{MNIST70K}, \textbf{MNIST2M} and \textbf{CIFAR-10}. Both PDBSCAN and \dbsc{} are run in 1 core per MPI process. The timings include the \knn{}/\enn{} construction.}
\label{table:MNIST} 
\setlength{\tabcolsep}{1pt} 
\begin{tabularx}{\columnwidth}{@{}l|C|CC|CC|CC}
\toprule %
  Dataset & \#cores &$T$  &NMI $\uparrow$ &$T$  &NMI $\uparrow$ &$T$  &NMI $\uparrow$\\
\hline
   \multicolumn{1}{c}{\cellcolor{black!20}\textbf{MNIST70K}} &\multicolumn{1}{c}{\cellcolor{black!20}$(\epsilon,M)$}   &\multicolumn{2}{c}{\cellcolor{black!20}$(1.20,50)$}    &\multicolumn{2}{c}{\cellcolor{black!20}$(1.20, 100)$} &\multicolumn{2}{c}{\cellcolor{black!20}$(1.39,100)$}\\
      \multicolumn{1}{c}{\cellcolor{black!20}\#} &\multicolumn{1}{c}{}   &\multicolumn{2}{c}{\labelText{\#1}{test:1}}    &\multicolumn{2}{c}{\labelText{\#2}{test:2}}  &\multicolumn{2}{c}{\labelText{\#3}{test:3}} \\\hline
    \multirow{2}{*}{PDBSCAN-OMP} &8 &469 &0.24 &469 &0.26 &515 &0.10\\
 &56 &82 &0.24 &83 &0.26 &94 &0.10\\
    \multirow{2}{*}{$k$NN-DBSCAN} &8 &62 &0.27 &80 &0.26 &80 &0.10\\
 &56 &11 &0.27 &14 &0.26 &14 &0.10\\\hline
   \multicolumn{1}{c}{\cellcolor{black!20}\textbf{MNIST2M}} &\multicolumn{1}{c}{\cellcolor{black!20}$(\epsilon,M)$}   &\multicolumn{2}{c}{\cellcolor{black!20}$(28.6,100)$}    &\multicolumn{2}{c}{\cellcolor{black!20}$(28.6, 200)$} &\multicolumn{2}{c}{\cellcolor{black!20}$(50.0,200)$}\\
    \multicolumn{1}{c}{\cellcolor{black!20}\#} &\multicolumn{1}{c}{}   &\multicolumn{2}{c}{\labelText{\#4}{test:4}}    &\multicolumn{2}{c}{\labelText{\#5}{test:5}}  &\multicolumn{2}{c}{\labelText{\#6}{test:6}} \\\hline
   \multirow{2}{*}{PDBSCAN} &16 &190 &0.93 &180 &0.93 &703 &0.94\\
&64 &57 &0.93 &55 &0.93  &246 &0.94\\
    \multirow{2}{*}{$k$NN-DBSCAN} &16 &64 &0.93 &80 &0.93 &80 &0.94\\
&64 &16 &0.93 &20 &0.93 &20 &0.94\\\hline
   \multicolumn{1}{c}{\cellcolor{black!20}\textbf{CIFAR-10}} &\multicolumn{1}{c}{\cellcolor{black!20}$(\epsilon,M)$}   &\multicolumn{2}{c}{\cellcolor{black!20}$(0.93,100)$}    &\multicolumn{2}{c}{\cellcolor{black!20}$(0.93, 200)$} &\multicolumn{2}{c}{\cellcolor{black!20}$(1.27,200)$}\\
    \multicolumn{1}{c}{\cellcolor{black!20}\#} &\multicolumn{1}{c}{}   &\multicolumn{2}{c}{\labelText{\#7}{test:7}}    &\multicolumn{2}{c}{\labelText{\#8}{test:8}}  &\multicolumn{2}{c}{\labelText{\#9}{test:9}} \\\hline
   \multirow{2}{*}{PDBSCAN} &16 &398 &0.060 &384 &0.059 &537 &0.022\\
&64 &107 &0.060 &111 &0.059  &156 &0.022\\
    \multirow{2}{*}{$k$NN-DBSCAN} &16 &115 &0.060 &84 &0.059 &84 &0.022\\
&64 &30 &0.060 &24 &0.059 &24 &0.022\\\bottomrule
\end{tabularx}
\end{table}

In this section, we compare \dbsc{} with PDBSCAN \cite{patwary2012,pdbscan-patwary12} using three real-world datasets (\textbf{MNIST70K}, \textbf{MNIST2M}, and \textbf{CIFAR-10}) and two synthetic datasets (\textbf{Uniform1S} and \textbf{Uniform2S}). PDBSCAN has two versions: a single-node multithreaded version using OpenMP and an MPI version (which does not support hybrid parallelism). Although \dbsc{} supports hybrid parallelism, we use one MPI process per core for a fair comparison with PDBSCAN. Our reported timings include the time taken for both the nearest neighbor graph construction and the clustering process. PDBSCAN constructs an exact \enn{} internally using a binary KD-tree. We vary the $M$ and $\epsilon$ parameters to study their effect on the timings and clustering accuracy. Note that the $\epsilon$ values used in the experiments reported in this section are relative values of $\epsilon/\widetilde{r}_2$, where $\widetilde{r}_2$ is the median distance to the second nearest neighbor.

\begin{figure}[t]
    \centering
\begin{tikzpicture}
        \node[inner sep=0pt] (a1) at (0,0) {\includegraphics[width=6cm]{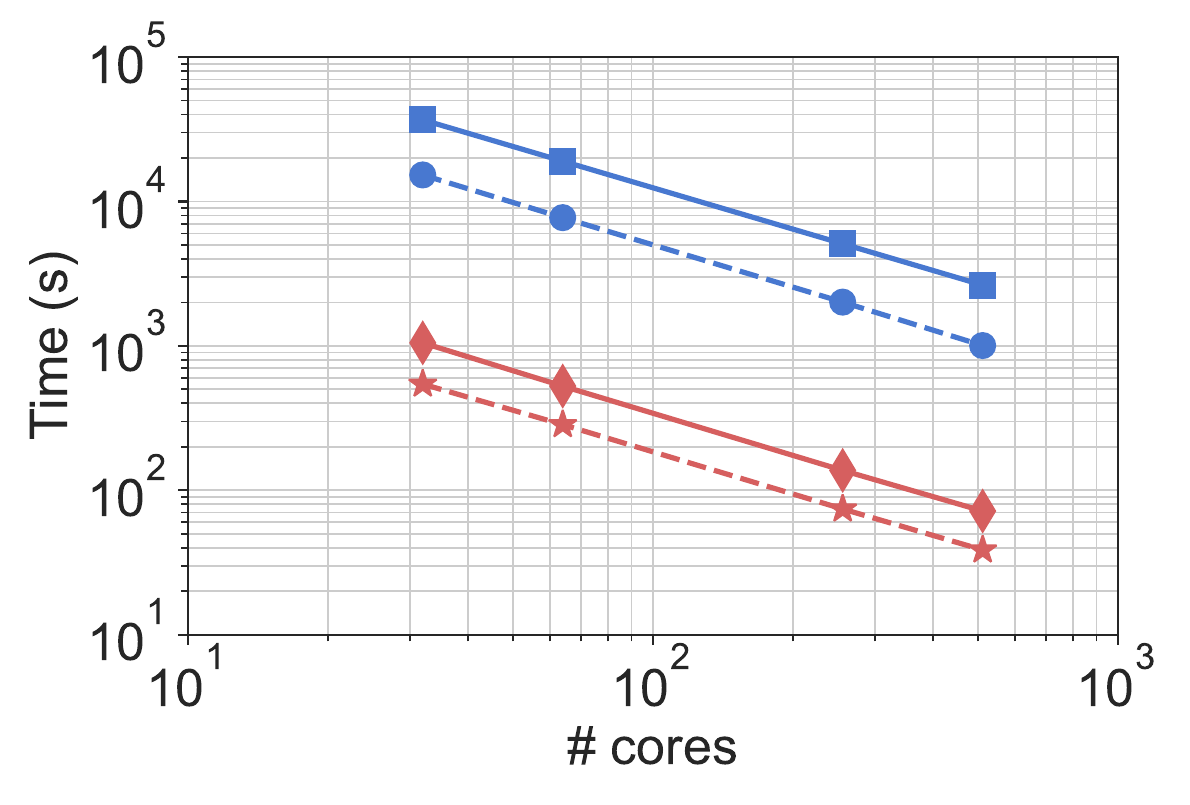}};
    \node[draw, inner sep=0pt] (a2) at (6,0) {\includegraphics[width=4cm]{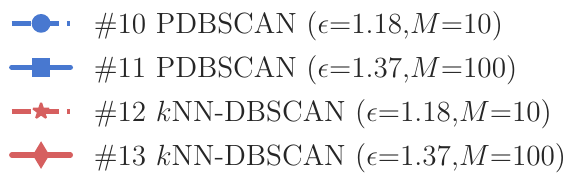}};
\end{tikzpicture}
    \caption{Strong scaling comparison between PDBSCAN and $k$NN-DBSCAN on \textbf{Uniform1S} with 2 MPI processes per node. The input parameters $\epsilon$ and $M$ for each test are indicated in the legend for each test.}
    \label{fig:Uniform1s}
\end{figure}

We label all comparison experiments  from \#1 to \#15. The datasets are summarized in \Cref{table:datasets}.  In \Cref{table:MNIST}, {we report wall-clock time and Normalized Mutual Information (NMI) of the clustering results for real-world datasets given different number of cores and hyperparameters. NMI is a commonly used score to evaluate the quality of clustering results, which is defined in~\Cref{sec:NMI}.} The MPI version of PDBSCAN fails on \textbf{MNIST70K} for unknown reasons, so we report timings using the OpenMP variant (denoted as PDBSCAN-OMP).   In \Cref{fig:Uniform1s} and \Cref{table:Uniform2s}, we present comparison results between \dbsc{} and PDBSCAN on \textbf{Uniform1S} and \textbf{Uniform2S}, respectively.

\textbf{Accuracy.}
In tests of~\Cref{table:MNIST}, the clustering quality of \dbsc{} is almost identical to PDBSCAN. NMI values of \dbsc{} and PDBSCAN are the same in runs \#\nameref{test:2}\textendash\#\nameref{test:9}. In \#\nameref{test:1}, NMI of \dbsc{} (0.27) is \textit{higher} than PDBSCAN (0.24). In fact, PDBSCAN generates 6 clusters in this run, while \dbsc{} generates 7 clusters. This demonstrates what we derive in Theorem \ref{thm:subsets}, i.e. the cluster number of \dbsc{} is no less than the cluster number of DBSCAN.  For dataset \textbf{Uniform2S}, points are uniformly distributed on the surface of two spheres in 10 dimensions. \Cref{table:Uniform2s}(a) presents the median distances of both surfaces.  To recover both clusters, we select $\epsilon = \widetilde{r}_{100}/\widetilde{r}_2$ of the \textit{outer} surface. PDBSCAN fails in \nameref{test:14} 
 since it runs out of memory caused by the $\epsilon$-range searches for the  points on the \textit{inner} surface. In contrast, \dbsc{} works fine in \nameref{test:15} because it uses a $k$NN graph, which is more robust to density variations compared to using an $\epsilon$-range graph.

\begin{figure}[t!]
    \centering
\includegraphics[totalheight=5cm]{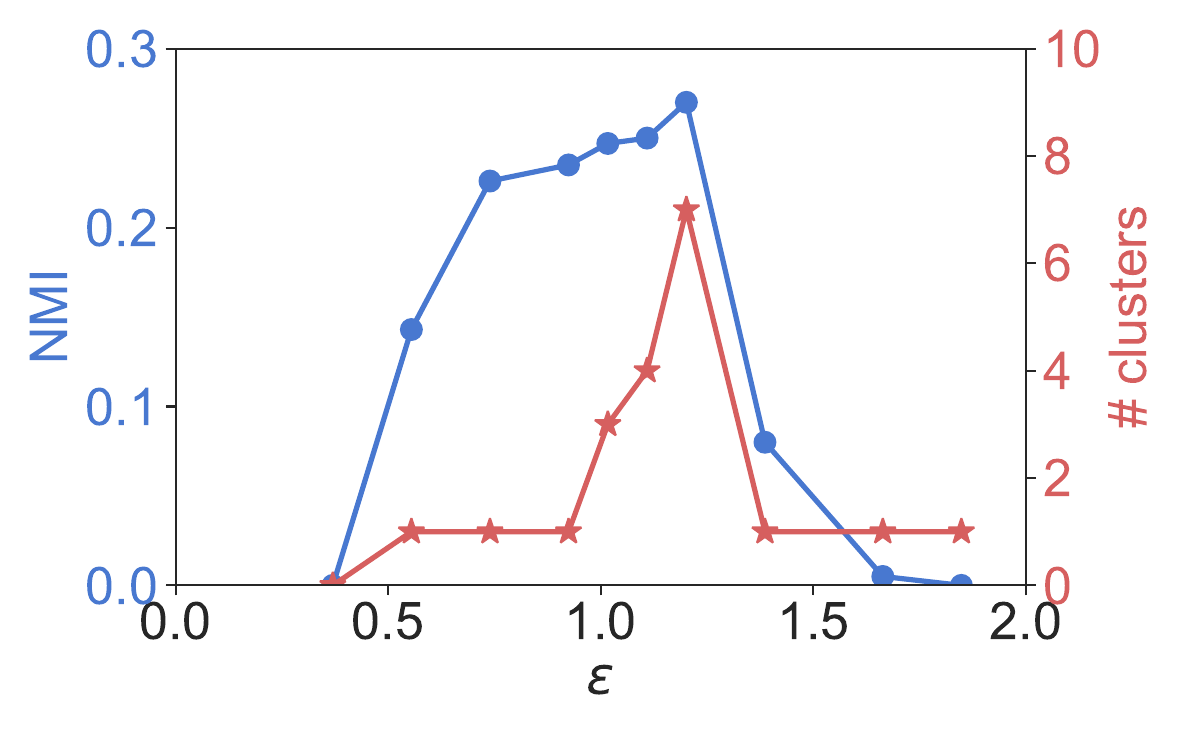}
    \caption{Clustering results (NMI and number of clusters) of $k$NN-DBSCAN on \textbf{MNIST70K} with $M=50$ and different values of $\epsilon$.}
    \label{fig:eps_mnist}
\end{figure}

\textbf{Efficiency.}
It is evident from ~\Cref{table:MNIST}, \Cref{fig:Uniform1s} and \Cref{table:Uniform2s} that $k$NN-DBSCAN is in general much faster than PDBSCAN. For instance, $k$NN-DBSCAN is 37 times faster than PDBSCAN on \Cref{fig:Uniform1s} when $\epsilon = 1.18$ and $M = 10$ (as seen in experiments \#10 and \#12). In~\Cref{fig:eps_mnist}, we report \dbsc{}'s NMI and the number of clusters as a function of $\epsilon$ for \textbf{MNIST70K}; the best NMI (0.27) with 7 clusters is achieved at $\epsilon = 1.20$, corresponding to \nameref{test:1}. We can observe that finding the optimal $\epsilon$ does \textbf{not} require recomputation of the \knn{} for $k$NN-DBSCAN, unlike PDBSCAN. Additionally, using a looser GOFMM accuracy can further reduce the timings for \dbsc{}. In fact, for most datasets, reducing the accuracy to 75\% does not impact the clustering quality.

\begin{table}[!tb]
\footnotesize
    \caption{Comparison between PDBSCAN and $k$NN-DBSCAN on \textbf{Uniform2S}. (a) $\widetilde{r}_2$ and $\widetilde{r}_{100}$ represent the median distances to the 2nd and 100th nearest neighbors, respectively. (b) For all tests, $M=100$ and $\epsilon$ is set to $\widetilde{r}_{100}/\widetilde{r}_2$ of the outer surface.  }
    \begin{subtable}{.5\linewidth}
      \centering
        \caption{}
        \label{table:Uniform2s}
\begin{tabular}{c|c|c}
    \toprule
    sphere & $\widetilde{r}_2$ &$\widetilde{r}_{100}$\\
    \midrule
    inner & 0.03& 0.046 \\
    outer &0.23 & 0.41\\
    \bottomrule
\end{tabular}
    \end{subtable}%
    \begin{subtable}{.5\linewidth}
      \centering
        \caption{}
\begin{tabular}{c|c|c}
    \toprule
    \# cores & PDBSCAN &$k$NN-DBSCAN\\
    \# & \labelText{\#14}{test:14} & \labelText{\#15}{test:15}\\
    \midrule
    8 & Fail& 1,425s \\
    64 &Fail & 199s\\
    512 & Fail & 49s\\
    \bottomrule
\end{tabular}
    \end{subtable} 
\end{table}

%% file: s4_conclusions.tex
We presented \dbsc{}, an algorithm based on  $k$-nearest-neighbor graphs, proved its correctness,  and demonstrated scalability both in the problem size and the dimension.  In our implementation the most expensive part of the algorithm is the construction of \knn{}, which can be considered as a preprocessing step for each dataset.

{\bf Limitations and future directions:}
\dbsc{} is not formally equivalent to DBSCAN. For certain parameter values they result in similar clustering and in our test the results were indistinguishable. From a theoretical perspective, one open question is the effect of \knn{} accuracy on clustering. But we remark that, since clustering is very fast, we could perform clustering at every \knn{} iteration and terminate when the NMI between subsequent clusterings has converged to some specified accuracy. In this way, the effect of inexactness in \knn{} can be fully controlled. From a practical perspective, an interesting direction is to extend our approach to the OPTICS algorithm~\cite{ankerst-e99opticsdbscan} that addresses a limitation of DBSCAN. The latter assumes that $\epsilon$ is fixed in the whole dataset.  Another possible avenue for improvement is to consider GPU acceleration.

%% file: appendix/proofs.tex
\section{Proofs}\label{sec:proof}

\subsection{Proof of \Cref{lemma:points_knn}}

\begin{proof}
    (i) From \defref{def:points}, $p\in\core$ $\Longleftrightarrow$ $|\mathcal{N}_\epsilon(p)| \geq M $ $\Longleftrightarrow$ $\max\limits_{q\in\mathcal{N}_M(p)} d(p,q)\leq \epsilon$.\\
     \indent (ii) ($\Longrightarrow$) If $p \notin \core$ and $p\in\border$, then $\exists \,q \in \core$, s.t. $d(p,q)\leq \epsilon$. We only need to show $q\in \mathcal{N}_M(p)$. Suppose to the contrary that $q\notin \mathcal{N}_M(p)$, then the distance from $q$ to $p$ is further than the $m$th nearest neighbor of $p$. This indicates that $\max\limits_{o\in\mathcal{N}_M(p)} d(p,o) \leq d(p,q) \leq \epsilon$, i.e. $p\in \core$, a contradiction.\\
     \indent ($\Longleftarrow$) trivial.\\
     \indent (iii) This follows directly from the definition of noise point in \defref{def:points}.
\end{proof}

\subsection{Proof of \Cref{lemma:equivalence}}
\begin{proof}
\begin{enumerate}[label=(\roman*)]
    \item  $M\mh\text{reachable}$ is reflexive: $p\in \core$, $p\in \mathcal{N}_M(p)$ $\Longrightarrow$ $p\xrightharpoonup{M} p$;\\
    \indent symmetric: $p, \,q \in \core$, $p\xrightharpoonup{M} q\Longrightarrow$ $\exists \, p_1, p_2, ..., p_n \in \core$, s.t. $p\xrightarrow{M} p_1$, $p_1\xrightarrow{M} p_2$, ..., $p_n\xrightarrow{M} q$. Since direct $M\mh\text{reachable}$ is symmetric for core points from \defref{def:k_direct}, $q \xrightarrow{M}p_n$, $p_n \xrightarrow{M}p_{n-1}$, ..., $p_1 \xrightarrow{M} p$ $\Longrightarrow q\xrightharpoonup{M} p$;\\
    \indent transitive: $p,\,q,\,r \in \core$, $p\xrightharpoonup{M} q$ and $q\xrightharpoonup{M} r$ $\Longrightarrow p\xrightharpoonup{M} r$.\\
    \indent Thus $M\mh\text{reachable}$ is a equivalence relation for all core points.
    \item Let $p, \, q\in \core $. If $p\xrightharpoonup{M} q$, then $p \xleftrightarrow{M} q$ since $p$ is also $M\mh\text{reachable}$ from $p$ itself. If $p \xleftrightarrow{M} q$, then $\exists \, o\in \core$, s.t. $o\xrightharpoonup{M} p$ and $o \xrightharpoonup{M}q$. From (i),  $M\mh\text{reachable}$ is symmetric and transitive, then $p\xrightharpoonup{M} o$, $o\xrightharpoonup{M} q$ and thus $p\xrightharpoonup{M} q$. 
\end{enumerate}
\end{proof}

\subsection{Proof of \Cref{lemma:relations}}

\begin{proof}
    (i) $p \xrightarrow{M} q$ $\Longrightarrow q\in \mathcal{N}_M(p)$ or $p \in \mathcal{N}_M(q)$ $\Longrightarrow d(p,q) \leq \epsilon$ $\Longrightarrow q \in \mathcal{N}_\epsilon(p)$ $\Longrightarrow p \directRadireach q$.\\
    \indent (ii) $p \xrightharpoonup{M}q$ $\Longrightarrow$ $\exists \, p_1, p_2, ..., p_n \in \core$, s.t. $p\xrightarrow{M} p_1$, $p_1\xrightarrow{M} p_2$, ..., $p_n\xrightarrow{M} q$. \\
    \indent From (i), $p\directRadireach p_1$, $p_1\directRadireach p_2$, ..., $p_n\directRadireach q$ $\Longrightarrow p \radireach q$.\\
    \indent (iii) This property directly follows from (ii) since reachable relations are equivalent to the connected relations (for core points) in both DBSCAN and \dbsc{}.
\end{proof}

\subsection{Proof of Theorem~\ref{thm:mst}}

\begin{proof}
    (i) Pick arbitrary $p,\,q\in \core$, $p$ and $q$ belong to a same MST cluster w.r.t $G_{\epsilon,\text{core}}$. \\
    \indent $\Longleftrightarrow$ $p$ and $q$ are in a same MST subtree, i.e. $\exists \, p_1, p_2, ..., p_n \in \core$, s.t. there are edges connecting $p$ and $p_1$, $p_1$ and $p_2$, ..., $p_n$ and $q$.\\
    \indent $\Longleftrightarrow$ $\exists \, p_1, p_2, ..., p_n \in \core$, s.t. $p_1\in\mathcal{N}_\epsilon(p)$, $p_2\in\mathcal{N}_\epsilon(p_1)$, ..., $q\in\mathcal{N}_\epsilon(p_n)$, i.e. $p\directRadireach p_1$, $p_1 \directRadireach p_2$, ..., $p_n \directRadireach q$.\\
    \indent $\Longleftrightarrow$ $p\radireach q$ $\Longleftrightarrow$ $p$ and $q$ belong to a same DBSCAN cluster. \\
    \indent (ii) Pick arbitrary $p,\,q\in \core$, $p$ and $q$ belong to a same MST cluster w.r.t $G_{M,\text{core}}$. \\
    \indent $\Longleftrightarrow$ $\exists \, p_1, p_2, ..., p_n \in \core$, s.t. there are edges connecting $p$ and $p_1$, $p_1$ and $p_2$, ..., $p_n$ and $q$.\\
    \indent $\Longleftrightarrow$ $\exists \, p_1, p_2, ..., p_n \in \core$, s.t. $p_1\in\mathcal{N}_M(p)\vee p\in\mathcal{N}_M(p_1)$, $p_1\in\mathcal{N}_M(p_2) \vee p_2\in\mathcal{N}_M(p_1)$, ..., $p_n\in\mathcal{N}_M(q) \vee q\in\mathcal{N}_M(p_n)$, i.e. $p \xrightarrow{M} p_1$, $p_1 \xrightarrow{M}  p_2$, ..., $p_n \xrightarrow{M}  q$.\\
    \indent $\Longleftrightarrow$ $p\xrightharpoonup{M}q$ $\Longleftrightarrow$ $p$ and $q$ belong to a same \dbsc{} cluster. \\
\end{proof}

\subsection{Proof of Theorem~\ref{thm:inexactmst}}

\begin{proof}
    (i) Given  $G_{\epsilon,\text{core}} = (\core ,\, E)$, it's enough to show that for arbitrary $p,\, q \in \core$, $p$ and $q$ belong to a same inexact MST cluster $\Longleftrightarrow$ $p\radireach q$. 
    
    ($\Longrightarrow$) If $p$ and $q$ belong to a same inexact MST subtree, then $\exists \, p_1, \,p_2$, ..., $p_n \in \core$, s.t. $p \directRadireach p_1$, $p_1 \directRadireach p_2$, ..., $p_n \directRadireach q$, i.e. $p \radireach q$. 
    
    ($\Longleftarrow$) If $p\radireach q$, then $\exists \, p_1, \,p_2$, ..., $p_n \in \core$, s.t. $p \directRadireach p_1$, $p_1 \directRadireach p_2$, ..., $p_n \directRadireach q$. We denote the super vertices of these points formed in the local step as $\widehat{p}$, $\widehat{p}_1$, ..., $\widehat{p}_n$ and $\widehat{q}$.
    
    If super vertices $\widehat{p}$ and $\widehat{q}$ are the same, points $p$ and $q$ are already in a same group. Since the second step of inexact MST construction can only combine but not divide these groups, $p$ and $q$ belong to a same inexact MST cluster eventually. 
    
    If $\widehat{p}$ and $\widehat{q}$ are different super vertices, we can choose a subsequence $\{ \widehat{p}_{k_i} \}_{i=0}^m$ from $\{\widehat{p},\, \widehat{p}_1, ..., \widehat{p}_n, \,\widehat{q}\}$ to satisfy the following conditions: 
    \begin{itemize}[noitemsep,topsep=0pt]
        \item {$\widehat{p}_{k_0} =p$, $\widehat{p}_{k_m} =q$, $1 \leq k_1 < k_2 < ... < k_{m-1} \leq n$.}
        \item $\widehat{p}_{k_i}$ and $\widehat{p}_{k_{i+1}}$ are different super vertices, $i=0,1,...,m-1$.
        \item $m$ is the maximal possible number of such sequence.
    \end{itemize}
    Then,  there are edges connecting $\widehat{p}$ 
   and $\widehat{p}_{k_1}$, $\widehat{p}_{k_1}$ and $\widehat{p}_{k_2}$, ..., $\widehat{p}_{k_m}$ and $\widehat{q}$. We claim that all such edges are \textit{cut} edges from $E_\text{cut}$. Indeed, the first local MST step guarantees that for every two different super vertices, there is \textit{no local} edge connecting them. Since every $\widehat{p}_i$ and $\widehat{p}_{i+1}$ of the newly formed sequence are different, edges connecting them can only be cut edges. This implies that $\widehat{p}$ and $\widehat{q}$ belong to a same cut MST subtree after the second step of inexact MST construction. Therefore, $p$ and $q$ are in the same inexact MST cluster.
   
  (ii) The proof follows a similar argument as (i).
\end{proof}

%% file: appendix/experiment_additional.tex
\section{Silhouette score and Normalized Mutual Information (NMI)}
\subsection{\textcolor{black}{Silhouette score}}\label{sec:ss}

\textcolor{black}{
The silhouette value is a measure of how similar an object is to its own cluster (cohesion) compared to other clusters (separation). For a data point $i$ in its cluster $C_I$, define $a(i)$ as the mean distance between  $i$ and all other data points in the same cluster, i.e.
\begin{align*}
    a(i) = \frac{1}{|C_I|- 1}\sum_{j\in C_I, i\neq j} d(i,j),
\end{align*}
define $b(i)$ be the smallest mean distance of  $i$ to all points in any other cluster, i.e.
\begin{align*}
    b(i) = \argmin_{J \neq I}\frac{1}{|C_J|} \sum_{j\in C_J}d(i,j).
\end{align*}}

\textcolor{black}{
The silhouette score of a point $i$ is defined as 
\begin{align*}
    s(i) =  \begin{cases}
      \frac{b(i) - a(i)}{\max\{a(i),b(i)\}} & \text{if }|C_I| >1\\
      0 & \text{if }|C_I| =1.
    \end{cases}    
\end{align*}
The silhouette score of the clustering results is defined as the average silhouette score across all data points. The silhouette score ranges from -1 to +1, where a high value indicates that the object is well matched to its own cluster and poorly matched to neighboring clusters.}

\subsection{NMI}\label{sec:NMI}
NMI is a score between $0$ and $1$ used to evaluate clustering quality (the higher the better). Given a set of points, let $Y$ be the true class labels, $C$ be the clustering labels, NMI is defined by
\begin{align*}
    NMI = \frac{2 I(Y;C)}{H(Y) + H(C)},
\end{align*}
where $H(\cdot)$ is the entropy, $I(Y;C)$ is the mutual information between $Y$ and $C$, which is given as
\begin{align*}
   I(Y;C) =  H(Y) - H(Y|C).
\end{align*}